\documentstyle[11pt,amsmath,amssymb,epsf,epic,cite]{article}
\numberwithin{equation}{section}

\newcommand{\be}{\begin{equation}}
\newcommand{\ee}{\end{equation}}
\newcommand{\bea}{\begin{eqnarray}}
\newcommand{\eea}{\end{eqnarray}}

\newcommand{\e}{{\rm e}}

\renewcommand{\d}{{\rm d}}

\newcommand{\grintl}{[\kern-.18em [}
\newcommand{\grintr}{]\kern-.18em ]}

\newcounter{resultcounter}[section]

\newtheorem{definition}[resultcounter]{Definition}

\def\bed{\begin{definition}}
\def\eed{\end{definition}}

\newcommand{\r}{{\rm R}}
\newcommand{\s}{{\rm S}}
\newcommand{\h}{{\cal H}}
\newcommand{\cx}{{\mathbb C}}
\newcommand{\rx}{{\mathbb R}}
\renewcommand{\i}{{\rm i}}

\newcommand{\fer}[1]{(\ref{#1})}
\newcommand{\scalprod}[2]{\left\langle {#1}, {#2}\right\rangle}
\newcommand{\bbbone}{\mathchoice {\rm 1\mskip-4mu l} {\rm 1\mskip-4mu l}
{\rm 1\mskip-4.5mu l} {\rm 1\mskip-5mu l}}

\begin{document}

\title{Evolution of Entanglement of Two Qubits
Interacting through Local and Collective Environments}

\author{M. Merkli\footnote{Department of Mathematics and Statistics,  Memorial
University of Newfoundland, St. John's, Newfoundland, Canada A1C
5S7.}\ \footnote{ Supported by NSERC under Discovery Grant 205247.
Email: merkli@mun.ca, http://www.math.mun.ca/merkli/ } \and  G.P.
Berman\footnote{Theoretical Division, MS B213, Los Alamos National
Laboratory, Los Alamos, NM 87545, USA. Email: gpb@lanl.gov. Work was
carried out under the auspices of  the NNSA of the U. S. DOE at LANL
under Contract No. DEAC52-06NA25396, and was partly supported by the
IARPA. } \and F. Borgonovi\footnote{Dipartimento di Matematica e
Fisica, Universit\`a Cattolica, via Musei 41, 25121 Brescia, Italy.
Email: fausto.borgonovi@unicatt.it, http://www.dmf.unicatt.it/$\sim$borgonov}\ \footnote{INFN, Sezione di
Pavia, Italy} \and K. Gebresellasie${}^*$\footnote{Supported by NSERC Discovery Grant 205247.}}
\date{\today}
\maketitle
\vspace{-.5cm}

\begin{abstract}
We analyze rigorously the dynamics of the entanglement between two
qubits which interact only through  collective and local
environments. Our approach is based on the resonance perturbation
theory which assumes a small interaction between the qubits and the
environments. The main advantage of our approach is that the
expressions for (i) characteristic time-scales, such as decoherence,
disentanglement, and relaxation, and (ii)  observables are not
limited by finite times. We introduce a new classification of
decoherence  times based on clustering of the reduced density matrix
elements. The characteristic dynamical properties such as creation
and decay of entanglement are examined. We also discuss possible
applications of our results for superconducting quantum computation
and quantum measurement technologies.
\end{abstract}

\thispagestyle{empty}
\setcounter{page}{1}
\setcounter{section}{1}
\setcounter{section}{0}

\section{Introduction}

Entanglement plays a very important role in quantum information
processes  \cite{yu1, wbps, AFOV, FLWDR} (see also references
therein). Even if  different parts of the quantum system (quantum
register) are initially disentangled, entanglement naturally appears
in the process of quantum protocols. This ``constructive
entanglement'' must be preserved during the time of quantum
information processing. On the other hand, the system generally
becomes entangled with the environment. This ``destructive
entanglement'' must be minimized in order to achieve a needed
fidelity of quantum algorithms. The importance of these effects
calls for  the development of rigorous mathematical tools for
analyzing the dynamics of entanglement and for controlling the
processes of constructive and destructive entanglement. Another
problem which is closely related to quantum information is quantum
measurement. Usually, for a qubit (quantum two-level system),
quantum measurements operate under the condition
$\hbar\omega>\!\!>k_BT$, where $T$ is the temperature,  $\omega$ is
the transition frequency, $\hbar$ is the Planck constant, and $k_B$
is the Boltzmann constant. This condition is widely used in
superconducting quantum computation, when $T\sim 10-20mK$ and
$\hbar\omega/k_B\sim 100-150 mK$. In this case, one can use
Josephson junctions (JJ) and superconducting quantum interference
devices (SQUIDs), both as qubits \cite{MSS, YHCCW, DM, Setal, Ketal}
and as spectrometers \cite{Cletal} measuring a spectrum of noise and
other important effects induced by the interaction with the
environment. Understanding the dynamical characteristics of
entanglement through the environment on a large time interval will
help to develop new technologies for measurements not only of
spectral properties,  but also of quantum correlations induced by
the environment.

In this paper, we develop a consistent perturbation theory of
quantum dynamics of entanglement which is valid for arbitrary times
$t\geq 0$. This is important in many real situations because (i) the
characteristic times which usually appear in quantum systems with
two and more qubits involve different time-scales, ranging from a
relatively fast decay of entanglement and different reduced density
matrix elements (decoherence) to possibly quite large relaxation
times, and  (ii) for not exactly solvable quantum Hamiltonians
(describing the energy exchange between the system and the
environment) one can only use a perturbative approach in order to
estimate the characteristic dynamical parameters of the system.
Note, that generally not only are the time-scales for decoherence
and entanglement different, but so are their functional
time-dependences. Indeed, usually the off-diagonal reduced density
matrix elements in the basis of the quantum register do not decay to
zero for large times, but remain at the level of $O(\lambda^2)$,
where $\lambda$ is a characteristic constant of interaction between
a qubit and an environment \cite{MSB1}. On the other hand,
entanglement has a different functional time dependence, and in many
cases decays to zero in finite time. Another problem which we
analyze in this paper is a well-known cut-off  procedure which one
must introduce for high frequencies of the environment in order to
have finite expressions for the interaction Hamiltonian between the
quantum register and the environment. Generally, this artificial
cut-off frequency enters all expressions in the theory for physical
parameters, including decay rates and dynamics of observables. At
the same time, one does not have this cut-off problem in real
experimental situations. So, it would be very desirable to develop a
regular theoretical approach to derive physical expressions which do
not include the cut-off parameter. We show that our approach allows
us to derive these cut-off independent expressions as the main terms
of the perturbation theory, which is of $O(\lambda^2)$. The cut-off
terms are included in the corrections of $O(\lambda^4)$. At the same
time, the low-frequency divergencies still remain in the theory, and
need additional conditions for their removal.

 We describe the characteristic dynamical properties of the
simplest quantum register which consists of two not directly
interacting qubits (effective spins), which interact with local and
collective environments. We introduce  a classification of the
decoherence times based on a partition of the reduced density matrix
elements in the energy basis into clusters.  This classification,
valid for general $N$-level systems coupled to reservoirs, is rather
important for dealing with quantum algorithms with large registers.
Indeed, in this case different orders of ``quantumness"  decay on
different time-scales. The classification of decoherence time-scales
which we suggest will help to separate environment-induced effects
which are important from the unimportant ones for performing a
specific quantum algorithm. We point out that all the populations
(diagonal of density matrix) always belong to the same cluster to
which is associated the relaxation time.

We present analytical and numerical results for decay and creation
of entanglement for both solvable (integrable, energy conserving)
and unsolvable (non-integrable, energy-exchange) models, and explain
the relations between them.

This paper is devoted to a physical and numerical discussion of the
dynamical resonance theory, and its application to the evolution of
entanglement. A detailed exposition of the resonance method can be
found in \cite{MSB1,mm}. As the mathematical details leading to
certain expressions used in the discussion presented in this paper
are rather lengthy, we report them separately in \cite{mm}.

\section{Model}
We consider two qubits $\s_1$ and $\s_2$, each one coupled to a local
reservoir, and both together coupled to a collective reservoir. The
Hamiltonian of the two qubits is
\begin{equation}
H_\s=B_1S^z_1+B_2S^z_2,
\label{n1}
\end{equation}
where $B_j=\hbar\omega_j/{2}$ are effective magnetic fields,
$\omega_j$ is the transition frequency, and  $S_j^z$ is the Pauli
spin operator of qubit $j$. The eigenvalues of $H_\s$ are
\begin{equation}
E_1=B_1+B_2,\ E_2=B_1-B_2,\ E_3=-B_1+B_2,\ E_4=-B_1-B_2,
\label{46}
\end{equation}
with corresponding eigenstates
\begin{equation}
\varPhi_1 = |++\rangle,\ \varPhi_2 = |+-\rangle,\  \varPhi_3 = |-+\rangle,\ \varPhi_4 = |--\rangle,\
\label{Sbasis}
\end{equation}
where $S^z|\pm\rangle =\pm|\pm\rangle$.
Each of the three reservoirs consists of free thermal bosons at
temperature $T=1/\beta>0$, with Hamiltonian
\begin{equation}
H_{\r_j} = \sum_k \hbar\omega_k a_{j,k}^\dagger a_{j,k},\qquad j=1,2,3.
\label{n2}
\end{equation}
The index $3$ labels the collective reservoir. The creation and
annihilation  operators satisfy $[a_{j,k},a_{j',k'}^\dagger] =
\delta_{j,j'}\delta_{k,k'}$. The interaction between each qubit and
each reservoir has two parts: an energy conserving and an energy
exchange one. The total Hamiltonian is
\begin{eqnarray}
H &=& H_\s +H_{\r_1}+H_{\r_2} + H_{\r_3} \label{n3}\\
&& +\left( \lambda_1 S^x_1 +\lambda_2 S^x_2\right)\otimes\phi_3(g)\label{n4}\\
&& +\left( \kappa_1 S^z_1 +\kappa_2 S^z_2\right)\otimes\phi_3(f)\label{n5}\\
&& + \,\mu_1 S^x_1\otimes\phi_1(g) +\mu_2 S^x_2\otimes\phi_2(g)\label{n6}\\
&& + \,\nu_1 S^z_1\otimes\phi_1(f) +\nu_2 S^z_2\otimes\phi_2(f)\label{n7}.
\end{eqnarray}
Here, $\phi_j(g)=\frac{1}{\sqrt 2}(a_j^\dagger(g) +a_j(g))$, with
\begin{equation}
a_j^\dagger(g)=\sum_k g_k a^\dagger_{j,k},\qquad  a(g)=\sum_k g^*_k a_{j,k}.
\label{n8}
\end{equation}
The  $\lambda, \kappa, \mu,\nu $ are the dimensionless coupling
constants. The collective  interaction is given by \fer{n4}
(energy-exchange) and \fer{n5} (energy conserving), the local
interactions are given by \fer{n6}, \fer{n7}. Also, $S_j^x$ is the
spin-flip operator (Pauli matrix) of qubit $j$. In the continuous
mode limit, $g_k$ becomes a function $g(k)$, $k\in\rx^3$. Our
approach is based on analytic spectral deformation methods
\cite{MSB1} and requires some analyticity of the form factors $f,g$.
Instead of presenting this condition we will work here with examples
satisfying the condition.
\begin{itemize}
\item[(A)] Let $r\geq 0$, $\Sigma\in S^2$ be the spherical coordinates
of $\rx^3$. The form factors $h=f,g$ (see \fer{n4}-\fer{n7}) are $h(r,\Sigma)
= r^p\e^{-r^m}h_1(\Sigma)$, with $p=-1/2+n$, $n=0,1,\ldots$ and $m=1,2$. Here,
$h_1$ is any angular function.
\end{itemize}
This family contains the usual physical form factors \cite{PSE}.  We
point out that we include an ultraviolet cutoff in the interaction
in order for the model to be well defined. (The minimal mathematical
condition for this is that $f(k), g(k)$ be square integrable over
$k\in\rx^3$.) However, as discussed in point 2. before equation
\fer{cutoff}, our approach yields expressions for decay and
relaxation rates which, to lowest order in the couplings between the
qubits and the reservoirs, do {\it not} depend on the ultraviolet
characteristics of the model.

\section{Evolution of qubits: resonance approximation}
\label{sectevol}

We take initial states where the qubits are not entangled with the
 reservoirs. Let $\rho_\s$ be an arbitrary initial state of the qubits,
 and let $\rho_{\r_j}$ be the thermal equilibrium state of reservoir $\r_j$.
 Let $\rho_\s(t)$ be the reduced density matrix of the two qubits at time $t$.
 The reduced density matrix elements in the energy basis are
\begin{eqnarray}
\lefteqn{
[\rho_\s(t)]_{mn}:=\scalprod{\varPhi_m}{\rho_\s(t)\varPhi_n}}\nonumber\\
&& ={\rm Tr}_{\r_1+\r_2+\r_3}\left[ \rho_\s\otimes\rho_{\r_1}\otimes
\rho_{\r_2}\otimes\rho_{\r_3}\ \e^{-\i tH/\hbar} |\varPhi_n\rangle\langle\varPhi_m| \,
\e^{\i tH/\hbar}\right],
\label{40}
\end{eqnarray}
where we take the trace over all reservoir degrees of freedom.
Under the non-interacting dynamics (all coupling parameters zero),
we have
\begin{equation}
[\rho(t)]_{mn} = \e^{\i t e_{mn}/\hbar} [\rho(0)]_{mn}, \label{30}
\end{equation}
where $e_{mn}=E_m-E_n$.

In the rest of the paper we use the dimensionless functions and
parameters. For this we introduce a characteristic frequency,
$\omega_0$, to be defined later, in Section 8, and the dimensionless
energies, temperature, frequencies and wave vectors of thermal
excitations, and time by setting
\begin{equation}
E_n^\prime =  E_n/(\hbar\omega_0),\qquad
f_k^\prime=f_k/(\hbar\omega_0), \qquad
g_k^\prime=g_k/(\hbar\omega_0),\qquad T^\prime=k_BT/(\hbar\omega_0),
\end{equation}
$$
\beta^\prime=1/T^\prime,\qquad
\omega_k^\prime=\omega_k/\omega_0,\qquad\vec{k}^\prime=c\vec{k}/\omega_0,\qquad
t^\prime = \omega_0 t,
$$
where $c$ is the speed of light. Below we omit index ``prime" in all
expressions.

As the interactions with the reservoirs are turned on (some of
$\kappa_j,\lambda_j,\mu_j,\nu_j$ nonzero), the dynamics \fer{30}
undergoes two qualitative changes.
\begin{itemize}
\item[1.] The ``Bohr frequencies''
\begin{equation}
e\in \{ E_k-E_l\ : E_k,E_l\in{\rm spec}(H_\s)\} \label{17}
\end{equation}
in the exponent of \fer{30} become {\it complex resonance energies},
$e\mapsto \varepsilon_e$, satisfying $\Im\varepsilon_e\geq 0$. If
$\Im\varepsilon_e >0$ then the corresponding density matrix elements
decay to zero (irreversibility).
 \item[2.] The matrix elements do not evolve independently any more.
 To lowest order in the couplings, all matrix elements
with $(m,n)$ belonging to a fixed energy difference $E_m-E_n$ will
evolve in a coupled manner. Thus to a given energy difference $e$,
\fer{17},  we associate the cluster of matrix element indexes
\begin{equation}
{\cal C}(e)=\{ (k,l)\ :\ E_k-E_l=e\}.
\label{32}
\end{equation}
\end{itemize}
\noindent Both effects are small if the coupling is small, and they
can be described by perturbation theory of energy  differences
\fer{17}. We view the latter as the eigenvalues of the {\it
Liouville operator}
\begin{equation}
L_\s = H_\s\otimes\bbbone_{\s} - \bbbone_{\s}\otimes H_\s,
\label{16'}
\end{equation}
acting on the doubled Hilbert space $\h_\s\otimes\h_\s$ (and ${\cal
H}_\s=\cx^2\otimes\cx^2$). The appearance of `complex energies' for
open systems is well known to  stem from passing from a Hamiltonian
dynamics to an effective non-Hamiltonian one by tracing out
reservoir degrees of freedom. The fact that independent clusters arise in the dynamics to lowest
order in the coupling can be understood heuristically as follows.
The interactions change the effective energy of the two qubits, i.e.
the basis in which the reduced density matrix is diagonal. Thus the
eigenbasis of $L_\s$ \fer{16'} is changed. However, to lowest order
in the perturbation, spectral subspaces with fixed $e\in{\rm
spec}(L_\s)$ {\it are left invariant} and stay orthogonal for
different unperturbed $e$. So matrix elements associated to ${\cal
C}(e)$ get mixed, but they do not mix with those in ${\cal C}(e')$,
$e\neq e'$.

\medskip
Let $e$ be an eigenvalue of $L_\s$ of multiplicity ${\rm mult}(e)$.
As the coupling parameters are turned on, there are generally many
distinct resonance energies bifurcating out of $e$. We denote them
by $\varepsilon_e^{(s)}$, where the parameter $s$ distinguishes different
resonance energies and varies between $s=1$ and
$s=\nu(e)$, where $\nu(e)$ is some number not exceeding ${\rm
mult}(e)$. We have a perturbation expansion
\begin{equation}
\varepsilon_e^{(s)} = e +\delta_e^{(s)} +O(\varkappa^4),
\label{u1}
\end{equation}
 where
\begin{equation}
\varkappa:= \max\{ |\kappa_j|, |\lambda_j|, |\mu_j|, |\nu_j|\ :\
j=1,2\} \label{12}
\end{equation}
and where $\delta_e^{(s)}=O(\varkappa^2)$ and $\Im\delta_e^{(s)}\geq 0$. The lowest order corrections $\delta_e^{(s)}$ are the eigenvalues of an explicit {\em level shift operator} $\Lambda_e$ (see
\cite{MSB1}), acting on the eigenspace of $L_\s$ associated to $e$. There are two
bases $\{\eta_e^{(s)}\}$ and $\{\widetilde\eta_e^{(s)}\}$ of the eigenspace, satisfying
\begin{equation}
\Lambda_e\eta_e^{(s)} = \delta_e^{(s)}\eta_e^{(s)}, \qquad [\Lambda_e]^*\widetilde\eta_e^{(s)} = {\delta_e^{(s)}}^*\ \widetilde\eta_e^{(s)}, \qquad \scalprod{\widetilde\eta_e^{(s)}}{\eta_e^{(s')}} =\delta_{s,s'}.
\label{19}
\end{equation}
We call the eigenvectors $\eta_e^{(s)}$ and $\widetilde\eta_e^{(s)}$ the `resonance vectors'.  We take interaction parameters ($f,g$ and the coupling constants) such that the following condition is satisfied.
\begin{itemize}
\item[(F)] There is complete splitting of resonances under perturbation at second order, i.e., all the $\delta_e^{(s)}$ are distinct for fixed $e$ and varying $s$.
\end{itemize}
This condition implies in particular that there are ${\rm mult}(e)$ distinct resonance energies $\varepsilon_e^{(s)}$, $s=1,\ldots,{\rm mult}(e)$ bifurcating out of $e$, so that in the above notation, $\nu(e)={\rm mult}(e)$.
Explicit evaluation of $\delta_e^{(s)}$ shows that condition (F) is satisfied for generic values of the interaction parameters (see also \fer{52}-\fer{56}).

\medskip
The following result is obtained from a detailed analysis of a
representation of the reduced dynamics given in \cite{MSB1}, and generalized to the present model with three reservoirs. The mathematical details are presented in \cite{mm}.

\medskip
{\bf Result on reduced dynamics.\ }
{\it
Suppo\-se that Conditions (A) and (F) hold.  There
is a constant $\varkappa_0>0$ such that if $\varkappa
<\varkappa_0$, then we have for all $t\geq 0$
\begin{equation}
[\rho_t]_{mn} =\sum_{(k,l)\in{\cal C}(E_m-E_n)}A_t(m,n;k,l)\  [\rho_0]_{kl} +O(\varkappa^2),
\label{42}
\end{equation}
where the remainder term is uniform in $t\geq 0$, and where the amplitudes $A_t$ satisfy the {\em Chapman-Kolmogorov} equation
\begin{equation}
A_{t+r}(m,n;k,l)= \sum_{(p,q)\in{\cal C}(E_m-E_n)}A_t(m,n;p,q)A_r(p,q;k,l),
\label{chko}
\end{equation}
for $t,r\geq 0$, with initial condition $A_0(m,n;k,l) = \delta_{m=k}\delta_{n=l}$ (Kronecker delta). Moreover, the amplitudes are given in terms of
the resonance vectors and resonance energies by
\begin{equation}
A_t(m,n;k,l) =  \sum_{s=1}^{{\rm mult}(E_n-E_m)} \e^{\i
t\varepsilon_{E_n-E_m}^{(s)}}
\scalprod{\varPhi_l\otimes\varPhi_k}{\eta_{E_n-E_m}^{(s)}}
\scalprod{\widetilde\eta_{E_n-E_m}^{(s)}}{\varPhi_n\otimes\varPhi_m}.
 \label{35}
\end{equation}
}

{\it Remark.\ } The upper bound $\varkappa_0$ satisfies $\varkappa^2_0\leq {\rm const.\,} T$, where $T$ is the temperature of the reservoirs, \cite{MSB1}.

We will call the first term on the r.h.s. of \fer{42} the {\bf resonance approximation} of the reduced density matrix dynamics.

\medskip
\noindent
{\bf Discussion.\ } 1. The result shows that to lowest
order in $\varkappa$, {\em and homogeneously in time}, the reduced
density matrix elements evolve in clusters. A cluster is determined
by indices in a fixed ${\cal C}(e)$. Within each cluster the
dynamics has the structure of a Markov process. Moreover, the
transition amplitudes of this process are given by the resonance
data. They can be calculated explicitly in concrete models. We have
therefore a simple approximation of the true dynamics, valid
homogeneously in time. This is an  {\it advantage} of the resonance
representation. A {\it limitation} is that this method cannot
describe the evolution of quantities (averages of observables) which
are of the order of the square of the coupling parameters, since the
error in the approximation is of the same order.\footnote{However, by
including higher order terms in the perturbation theory, one can
refine the resonance method and resolve processes of higher order in
the coupling.} An illustration of this limitation of the method is the large-time behaviour of off-diagonal matrix elements. Generically, all off-diagonals decay to a limit having the size $O(\varkappa^2)$, as $t\rightarrow\infty$ \cite{MSB1}. As soon as a matrix element is of order $O(\varkappa^2)$, the resonance approximation \fer{42} cannot resolve its dynamics, since it is of the same order as the remainder.

One of our goals is to describe the evolution of entanglement of the qubits (see sections \ref{sectentcre}, \ref{numres}). From the above explanations, it is clear that the resonance approximation is well suited to describe decay of initial entanglement of qubits (if the initial entanglement is much larger than $O(\varkappa^2)$). On the other hand, an initially unentangled qubit state will typically become entangled due to the interaction with reservoirs. It is expected that the entanglement created may be of the same order as the error in the approximation \fer{42}, and hence the question arises if it is possible to describe this process using the resonance approximation.
The answer is positive, as we show numerically in section \ref{numres}: indeed we see that the amount of entanglement created is {\it independent} of the coupling strength. (The effect of changing $\varkappa$ is to shift the time-dependent curve of entanglement along the time-axis.)

2. {\bf Cluster classification of density matrix
elements.\ } The main dynamics partitions the reduced dynamics into
independent clusters of jointly evolving matrix elements, according
to \fer{32}. Depending on the energy level distribution of the two
isolated qubits and on the interaction parameters, each cluster has
its associated decay rate. It is possible that some clusters decay
very quickly, while some others stay populated for much longer
times. The resonance dynamics furnishes us with a very concrete
recipe telling us which parts of the matrix elements disappear when. This reveals a pattern of where in the density matrix quantum
properties are lost at which time. The same feature holds for an
arbitrary $N$-level system coupled to reservoirs, \cite{MSB1}, and
notably for complex systems ($N>\!>1$). In particular, this approach
may prove useful in the analysis of feasibility of quantum
algorithms performed on $N$-qubit registers. We point out that that {\it the diagonal belongs always to a single cluster}, namely the one associated with $e=0$. If the energies of the $N$-level system are degenerate, then some off-diagonal matrix elements belong to the same cluster as the diagonal as well.

3. The sum in \fer{42}
alone, which is the main term in the expansion, preserves the
hermiticity but not the positivity of density matrices. In other
words, the matrix obtained from this sum may have negative
eigenvalues. Since by adding $O(\varkappa^2)$ we do get a true
density matrix, the mentioned negativity of eigenvalues can only be
of $O(\varkappa^2)$. This can cause complications if one tries to
calculate for instance the concurrence by using the main term in
\fer{42} alone. Indeed, concurrence is not defined in general for a
`density matrix' having negative eigenvalues. See also section \ref{numres},
Numerical Results.

4. It is well known that the time decay of matrix
elements is not exponential for all times. For example, for small
times it is quadratic in $t$ \cite{PSE}. How is this behaviour
compatible with the representation \fer{42}, \fer{35}, where only
exponential time factors $\e^{\i t\varepsilon}$ are present? The
answer is that {\it up to an error term of $O(\varkappa^2)$}, the
``overlap coefficients'' (scalar products in \fer{35}) mix the
exponentials in such a way as to produce the correct time behaviour.

5. Since the coupled system has an equilibrium state, one of the
resonances $\varepsilon_0^{(s)}$ is always zero \cite{MLSO}, we set
$\varepsilon_0^{(1)}=0$. The condition $\Im\varepsilon_e^{(s)}>0$
for all $e,s$ except $e=0$, $s=1$ is equivalent to the entire system
(qubits plus reservoirs) converging to its equilibrium state for
large times.

\medskip
As we have remarked above, the decay of matrix elements is not in
general exponential, but we can nevertheless represent it
(approximate to order $\varkappa^2$) in terms of superpositions of
exponentials, for all times $t\geq 0$. In regimes where the actual
dynamics has exponential decay, the rates coincide with those we
obtain from the resonance theory (large time dynamics, see Section \ref{sectcomp} and also
\cite{PSE,MSB1}). It is therefore reasonable to define the
thermalization rate by
$$
\gamma^{\rm th} = \min_{s\geq 2}\Im\varepsilon_0^{(s)}\geq 0
$$
and the {\it cluster decoherence rate} associated to ${\cal C}(e)$, $e\neq 0$, by
$$
\gamma^{{\rm dec}}_e =\min_{1\leq s  \leq {\rm mult}(e)} \Im\varepsilon_e^{(s)} \geq 0.
$$
The interpretation is that the cluster of matrix elements of the
true density  matrix associated to $e\neq 0$ decays to zero, modulo
an error term $O(\varkappa^2)$, at the rate $\gamma_e^{\rm dec}$,
and the cluster containing the diagonal approaches its equilibrium
(Gibbs) value, modulo an error term $O(\varkappa^2)$, at rate
$\gamma^{\rm th}$. If $\gamma$ is any of the above rates, then
$\tau=1/\gamma$ is the corresponding (thermalization, decoherence)
time.

It should be understood that characterizing the dynamcs via the cluster decoherence and relaxation times corresponds to a `coarse graining': matrix elements are grouped into clusters and the dynamics of clusters is effectively given by a decoherence (or the relaxation) time. Expression \fer{42} gives much more detail, it gives the dynamics of each single matrix element. The breakup of the density matrix into individually evolving clusters may be advantageous especially in complex systems, where instead of two qubits, one deals with $N$-qubit registers.

\medskip
\noindent {\bf Remark on the markovian property.\ } In the first
point discussed  after \fer{35}, we remark that within a cluster,
our approximate dynamics of matrix elements has the form of a Markov
process. In the theory of markovian master equations, one constructs
commonly an approximate dynamics given by a markovian quantum
dynamical semigroup, generated by a so-called Lindblad (or weak
coupling) generator \cite{BP}. Our representation is {\it not} in
Lindblad form (indeed, it is not even a positive map on density
matrices). To make the meaning of the markovian property of our
dynamics clear, we consider a fixed cluster $\cal C$ and denote the
associated pairs of indices by $(m_k,n_k)$, $k=1,\ldots, K$. Retaining only the main part in \fer{42}, and making use of
\fer{chko} we obtain for $t,s\geq 0$
\begin{equation}
\left[
\begin{array}{c}
{}[\rho_{t+s}]_{m_1 n_1}\\
\vdots\\
{}[\rho_{t+s}]_{m_K n_K}
\end{array}
\right]
= A_{\cal C}(t)
\left[
\begin{array}{c}
{}[\rho_{s}]_{m_1 n_1}\\
\vdots\\
{}[\rho_{s}]_{m_K n_K}
\end{array}
\right],
\label{clustermarkov}
\end{equation}
where $[A_{\cal C}(t)]_{m_jn_j, m_ln_l} =A_t(m_j,n_j;m_l,n_l)$, c.f.
\fer{35}. Thus the dynamics of the vector having  as components the
density matrix elements, has the semi-group property in the time
variable, with generator $G_{\cal C}:=\frac{\d}{\d t}A_{\cal C}(0)$,
\begin{equation}
\left[
\begin{array}{c}
{}[\rho_t]_{m_1 n_1}\\
\vdots\\
{}[\rho_t]_{m_K n_K}
\end{array}
\right]
= \e^{tG_{\cal C}}
\left[
\begin{array}{c}
{}[\rho_{0}]_{m_1 n_1}\\
\vdots\\
{}[\rho_{0}]_{m_K n_K}
\end{array}
\right].
\label{clustermarkov1}
\end{equation}
This is the meaning of the Markov property of the resonance dynamics.

While the fact that our resonance approximation is not in the form of the weak coupling limit (Lindblad) may represent disadvantages in certain
applications, it may also allow for a description of effects possibly
not visible in a markovian master equation approach. Based on results \cite{yu1, BLC}, one may believe that revival of entanglement is a non-markovian effect, in the sense that it is not detectable under the markovian master equation
dynamics (however, we are not aware of any demonstration of this result). Nevertheless, as we show in our numerical analysis below, the resonance approximation captures this effect (see Figure \ref{f1}). We may attempt to explain this as follows. Each cluster is a (indpendent) markov process with its own decay rate, and while some clusters may depopulate very quickly, the ones responsible for creating revival of entanglement may stay alive for much longer times, hence enabling that process. Clearly, on time-scales larger than the biggest decoherence time of all clusters, the matrix is (approximately) diagonal, and typically no revival of entanglement is possible any more.

\section{Explicit resonance data}
We consider the Hamiltonian $H_\s$, \fer{n2},  with parameters
$0<B_1<B_2$ s.t. $B_2/B_1\neq 2$. This assumption is a
non-degeneracy condition which is not essential for the
applicability of our method (but lightens the exposition). The
eigenvalues of $H_\s$ are given by \fer{46} and the spectrum of
$L_\s$ is $\{e_1,\pm e_2,\pm e_3,\pm e_4,\pm e_5\}$, with
non-negative eigenvalues
\begin{equation}
e_1=0,\  e_2=2B_1,\  e_3=2B_2,\  e_4=2(B_2-B_1),\  e_5=2(B_1+B_2),
\label{45}
\end{equation}
having multiplicities $m_1=4$, $m_2=m_3=2$, $m_4=m_5=1$,
respectively. According to \fer{45}, the grouping of jointly
evolving elements of the density matrix above and on the diagonal
is given by\footnote{Since the density matrix is hermitian, it
suffices to know the evolution of the elements on and above the
diagonal.}
\begin{eqnarray}
{\cal C}_1 &:=&{\cal C}(e_1) =\{(1,1), (2,2), (3,3), (4,4)\}\label{47}\\
{\cal C}_2 &:=&{\cal C}(e_2) = \{ (1,3), (2,4)\}\label{48}\\
{\cal C}_3 &:=&{\cal C}(e_3) = \{ (1,2), (3,4)\}\label{49}\\
{\cal C}_4 &:=&{\cal C}(-e_4) = \{ (2,3)\}\label{50}\\
{\cal C}_5 &:=&{\cal C}(e_5) = \{ (1,4)\}\label{51}
\end{eqnarray}
There are five clusters of jointly evolving elements (on and above the diagonal). One cluster is the diagonal, represented by ${\cal C}_1$.
{} For $x>0$ and $h\in L^2(\rx^3,\d^3k)$ we define
\begin{equation}
\sigma_h(x) = 4 \pi x^2 \coth(\beta x) \int_{S^2} |h(2x,\Sigma)|^2\d\Sigma
\label{57}
\end{equation}
(spherical coordinates) and for $x=0$ we set
\begin{equation}
\sigma_h(0) = 4 \pi \lim_{x\downarrow 0} x^2 \coth(\beta x) \int_{S^2} |h(2x,\Sigma)|^2\d\Sigma.
\label{58}
\end{equation}
Furthermore, let
\begin{eqnarray}
Y_2 &=&  \big| \Im \left[ 4\kappa_1^2\kappa_2^2 r^2 -\i (\lambda_2^2+\mu_2^2)^2 \sigma^2_g(B_2) -4\i \kappa_1\kappa_2  \ (\lambda_2^2+\mu_2^2) \ r r_2'  \right]^{1/2}\big|, \label{69}\\
Y_3 &=&  \big| \Im \left[ 4\kappa_1^2\kappa_2^2 r^2 -\i (\lambda_1^2+\mu_1^2)^2 \sigma^2_g(B_1) -4\i \kappa_1\kappa_2  \ (\lambda_1^2+\mu_1^2) \ r r_1'  \right]^{1/2}\big|,\label{70}
\end{eqnarray}
(principal value square root with branch cut on negative real axis) where
\begin{equation}
r={\rm P.V.}\int_{\rx^3} \frac{|f|^2}{|k|}\d^3k,\qquad r_j' = 4B_j^2\int_{S^2}|g(2 B_j,\Sigma)|^2 \d\Sigma.
\label{pvr}
\end{equation}
The following results are obtained by an explicit calculation of level shift operators. Details are presented in \cite{mm}.

\medskip
{\bf Result on decoherence and thermalization rates.}
{\it
The thermalization and decoherence rates are given by
\begin{eqnarray}
\gamma^{\rm th} &=& \min_{j=1,2}\left\{ (\lambda_j^2+\mu_j^2)\sigma_g(B_j)\right\}+O(\varkappa^4) \label{52}\\
\gamma_2^{\rm dec} &=& \textstyle\frac 12 (\lambda^2_1+\mu_1^2)\sigma_g(B_1) + \textstyle\frac 12 (\lambda^2_2+\mu_2^2)\sigma_g(B_2) \nonumber\\
&& - Y_2 +(\kappa_1^2+\nu^2_1)\sigma_f(0) +O(\varkappa^4)\label{53}\\
\gamma_3^{\rm dec} &=& \textstyle\frac 12 (\lambda^2_1+\mu_1^2)\sigma_g(B_1) + \textstyle\frac 12 (\lambda^2_2+\mu_2^2)\sigma_g(B_2) \nonumber\\
&& - Y_3 +(\kappa_2^2+\nu^2_2)\sigma_f(0) +O(\varkappa^4)\label{54}\\
\gamma_4^{\rm dec} &=& (\lambda^2_1+\mu_1^2)\sigma_g(B_1) + (\lambda^2_2+\mu_2^2)\sigma_g(B_2) \nonumber\\
&& +\left[(\kappa_1-\kappa_2)^2 +\nu_1^2+\nu_2^2\right]\sigma_f(0) +O(\varkappa^4) \label{55}\\
\gamma_5^{\rm dec} &=& (\lambda^2_1+\mu_1^2)\sigma_g(B_1) + (\lambda^2_2+\mu_2^2)\sigma_g(B_2) \nonumber\\
&& +\left[(\kappa_1+\kappa_2)^2 +\nu_1^2+\nu_2^2\right]\sigma_f(0)+O(\varkappa^4) \label{56}
\end{eqnarray}
}

\noindent
{\bf Discussion.\ } 1. The thermalization rate depends on energy-exchange parameters $\lambda_j$, $\mu_j$ only. This is natural since an energy-conserving dynamics leaves the populations constant. If the interaction is purely energy-exchanging ($\kappa_j=\nu_j=0$), then all the rates depend {\it symmetrically} on the local and collective interactions, through $\lambda_j^2+\mu_j^2$. However, for purely energy-conserving interactions ($\lambda_j=\mu_j=0$) the rates are not symmetrical in the local and collective terms. (E.g. $\gamma^{\rm dec}_4$ depends only on local interaction if $\kappa_1=\kappa_2$.) The terms $Y_2$, $Y_3$ are complicated nonlinear combinations of exchange and conserving terms. This shows that effect of the energy exchange and conserving interactions are correlated.

2. We see from \fer{57}, \fer{58} that the leading orders of the rates \fer{52}-\fer{56} do not depend on an ultraviolet features of the form factors $f,g$. (However, $\sigma_{f,g}(0)$ depends on the infrared behaviour.) The coupling constants, e.g. $\lambda_j^2$ in \fer{52} multiply $\sigma_g(B_j)$, i.e., the rates involve quantities like (see \fer{57})
\begin{equation}
\pi \lambda_j^2\int_{\rx^3} \coth\big(\beta |k|/2\big)\  \big|g(|k|,\Sigma)\big|^2 \ \delta^{(1)}(|k|-2B_j) \ \d^3k.
\label{cutoff}
\end{equation}
The one-dimensional Dirac delta function appears due to energy
conservation of processes of order $\varkappa^2$, and $2B_j$ is (one
of) the Bohr frequencies of a qubit. Thus energy conservation
chooses the evaluation of the form factors at finite momenta and
thus an ultraviolet cutoff is not visible in these terms.
Nevertheless, we do not know how to control the error terms
$O(\varkappa^4)$ in \fer{52}-\fer{56} homogeneously in the cutoff.

3. The case of a single qubit interacting with a thermal bose gas has been extensively studied, and decoherence and thermalization rates for the spin-boson system have been found using different techniques, \cite{Leggett, SMS, Weiss}. We recover the spin-boson model by setting all our couplings in \fer{n3}-\fer{n7} to zero, except for $\lambda_1=\kappa_1\equiv\lambda$, and setting $f=g$. In this case, the spectral density $J(\omega)$ of the reservoir is linked to our quantity \fer{57} by
$$
J(\omega) = \frac{\sigma_h(\omega/2)}{\coth(\beta\omega/2)}.
$$
The relaxation rate is
$$
\gamma^{\rm th}=\frac 12\pi\lambda^2\coth(\beta B) J(2B),
$$
where $2B$ is the transition frequency of qubit (in units where $\hbar=1$), see \fer{n1}. The decoherence rate is given by
$$
\gamma^{\rm dec} = \frac{\gamma^{\rm th}}{2} + \lambda^2\pi \sigma_h(0),
$$
where $\sigma_h(0)$ is the limit as $\omega\rightarrow 0$
of $\coth(\beta\omega)J(2\omega)$. These rates obtained with our resonance
method agree with those obtained in \cite{Leggett, SMS, Weiss}
by the standard Bloch-Redfield approximation.

\medskip
\noindent {\bf Remark on the limitations of the resonance approximation.\ } As mentioned in section \ref{sectevol}, the dynamics \fer{42} can only resolve the evolution of quantities larger than $O(\varkappa^2)$. For instance, assume that in an initial state of the two qubits, all off-diagonal density matrix elements are of the order of unity (relative to $\varkappa$). As time increases, the off-diagonal matrix elements decrease, and for times $t$ satisfying $\e^{-t\gamma^{\rm dec}_j} \leq O(\varkappa^2)$, the off-diagonal cluster ${\cal C}_j$ is of the same size $O(\varkappa^2)$ as the error in \fer{42}. Hence the evolution of this cluster can be followed accurately
by the resonance approximation for times $t< \ln(\varkappa^{-2})/\gamma^{\rm dec}_j\propto\frac{\ln(\varkappa^{-2})}{\varkappa^2 (1+T)}$, where $T$ is the temperature. Here, $T, \varkappa$ (and other parameters) are dimensionless. To describe the cluster in question for larger times, one has to push the perturbation theory to higher order in $\varkappa$. It is now clear that if a cluster is initially not populated, the resonance approximation does not give any information about the evolution of this cluster, other than saying that its elements will be $O(\varkappa^2)$ for all times.

Below we investigate analytically decay of entanglement (section \ref{disentsect}) and numerically creation of entanglement (section \ref{numres}). For the same reasons as just outlined, an analytical study of entanglement decay is possible if the initial entanglement is large compared to $O(\varkappa^2)$. However, the study of creation of entanglement is more subtle from this point of view, since one must detect the emergence of entanglement, presumably of order $O(\varkappa^2)$ only, starting from zero entanglement. We show in our numerical analysis that entanglement of size 0.3 is created {\it independently} of the value of $\varkappa$ (ranging from 0.01 to 1). We are thus sure that the resonance approximation does detect creation of entanglement, {\it even} if it may be of the same order of magnitude as the couplings. Whether this is correct for other quantities than entanglement is not clear, and so far, only numerical investigations seem to be able to give an answer. As an example where things can go wrong with the resonance approximation we mention that for small times, the approximate density matrix has {\it negative eigenvalues}. This makes the notion of concurrence of the approximate density matrix ill-defined for small times.

\section{Comparison between exact solution and resonance approximation: explicitly solvable model}
\label{sectcomp}

We consider the system with Hamiltonian \fer{n3}-\fer{n8} and $\lambda_1=\lambda_2=0$, $\mu_1=\mu_2=0$, $\kappa_1=\kappa_2=\kappa$ and $\nu_1=\nu_2=\nu$. This energy-conserving model can be solved explicitly \cite{PSE,MSB1} and has the {\bf exact solution}
\begin{equation}
 [\rho_t]_{mn} = [ \rho_0]_{mn}\  \e^{-\i t(E_m-E_n)}
\ \e^{\i\kappa^2 a_{mn} S(t)}\ \e^{ -[\kappa^2 b_{mn} +\nu^2 c_{mn}]\Gamma(t)}
\label{exact}
\end{equation}
where
$$
(a_{mn}) = \left[
\begin{array}{cccc}
0 & -4 & -4 & 0\\
4 & 0 & 0 & 4\\
4 & 0 & 0 & 4\\
0 & -4 & -4 & 0
\end{array}
\right],
(b_{mn}) = \left[
\begin{array}{cccc}
0 & 4 & 4 & 16\\
4 & 0 & 0 & 4\\
4 & 0 & 0 & 4\\
16 & 4 & 4 & 0
\end{array}
\right],
(c_{mn}) = \left[
\begin{array}{cccc}
0 & 4 & 4 & 8\\
4 & 0 & 8 & 4\\
4 & 8 & 0 & 4\\
8 & 4 & 4 & 0
\end{array}
\right]
$$
and
\begin{eqnarray}
S(t) & = & \frac 12 \int_{\rx^3} |f(k)|^2 \ \frac{|k|t - \sin(|k|t)}{|k|^2}\d^3k
\label{S}\\
\Gamma(t) &=& \int_{\rx^3} |f(k)|^2\coth(\beta |k|/2)\frac{\sin^2(|k|t/2)}{|k|^2}\d^3 k.
\label{Gamma}
\end{eqnarray}
On the other hand, the {\it main} contribution (the sum) in \fer{42}
yields the {\bf resonance approximation} to the true dynamics, given
by
\begin{eqnarray}
{}[\rho_t]_{mm} &\doteq& [\rho_0]_{mm} \mbox{\qquad $m=1,2,3,4$}\label{l1}\\
{}[\rho_t]_{1n} &\doteq& \e^{-\i t(E_1-E_n)}\ \e^{-2\i t\kappa^2 r} \e^{-t(\kappa^2+\nu^2)\sigma_f(0)}[\rho_0]_{1n} \qquad n=2,3\label{l2}\\
{}[\rho_t]_{14} &\doteq& \e^{-\i t(E_1-E_4)}\ \e^{-t(4\kappa^2+2\nu^2)\sigma_f(0)}[\rho_0]_{14}\label{l3}\\
{}[\rho_t]_{23} &\doteq& \e^{-\i t(E_2-E_3)}\ \e^{-2t \kappa^2 \sigma_f(0)}[\rho_0]_{23}\label{l4}\\
{}[\rho_t]_{m4} &\doteq& \e^{-\i t(E_m-E_4)}\ \e^{2\i t\kappa^2 r}\ \e^{-t(\kappa^2+\nu^2)\sigma_f(0)}[\rho_0]_{m4} \qquad m=2,3 \label{l5}
\end{eqnarray}
The dotted equality sign $\doteq$ signifies that the left side equals the right side modulo an error term $O(\kappa^2+\nu^2)$, homogeneously in $t\geq 0$.\footnote{To arrive at \fer{l1}-\fer{l5} one calculates the $A_t$ in \fer{42} explicitly, to second order in $\kappa$ and $\nu$. The details are given in \cite{mm}.}
Clearly the decoherence function $\Gamma(t)$ and the phase $S(t)$ are nonlinear in $t$ and depend on the ultraviolet behaviour of $f$. On the other hand, our resonance theory approach yields a representation of the dynamics in terms of a superposition of exponentially decaying factors. From \fer{exact} and \fer{l1}-\fer{l5} we see that the resonance approximation is obtained from the exact solution by making the replacements
\begin{eqnarray}
S(t) &\mapsto& \textstyle\frac{1}{2} r t,\label{mm30}\\
\Gamma(t) &\mapsto& \textstyle\frac{1}{4} \sigma_f(0) t.\label{mm31}
\end{eqnarray}
We emphasize again that, according to \fer{42}, the difference
between the  exact solution and the one given by the resonance
approximation is of the order $O(\kappa^2+\nu^2)$, homogeneously in
time, and where $O(\kappa^2+\nu^2)$ depends on the ultraviolet
behaviour of the couplings. This shows in particular that up to
errors of $O(\kappa^2+\nu^2)$, the dynamics of density matrix
elements is simply given by a phase change and a possibly decaying
exponential factor, both linear in time and entirely determined by
$r$ and $\sigma_f(0)$. Of course, the advantage of the resonance
approximation is that even for not exactly solvable models, we can
approximate the true (unknown) dynamics by an explicitly calculable
superposition of exponentials with exponents linear in time,
according to \fer{42}. Let us finally mention that one easily sees
that
$$
\lim_{t\rightarrow\infty}S(t)/t=r/2 \mbox{\quad  and\quad}  \lim_{t\rightarrow\infty}\Gamma(t)/t=\sigma_f(0)/4,
$$
so \fer{mm30} and \fer{mm31} may indicate that the resonance approximation is closer to the true dynamics for large times -- but nevertheless, our analysis proves that the two are close together ($O(\kappa^2+\nu^2)$) {\it homogeneously} in $t\geq 0$.

\section{Disentanglement}
\label{disentsect}

In this section we apply the resonance method to obtain estimates on
survival and death of entanglement under the full dynamics
\fer{n3}-\fer{n7} and for an initial state of the form
$\rho_\s\otimes\rho_{\r_1}\otimes \rho_{\r_2}\otimes\rho_{\r_3}$,
where $\rho_\s$ has nonzero entanglement and the reservoir initial
states are thermal, at fixed temperature $T=1/\beta>0$. Let $\rho$
be the density matrix of two qubits $1/2$. The {\it concurrence}
\cite{Wo, BDSW} is defined by
\begin{equation}
C(\rho) = \max\{ 0, \sqrt{\nu_1}-\big[\sqrt{\nu_2}+\sqrt{\nu_3}+\sqrt{\nu_4}\big]\},
\label{60}
\end{equation}
where $\nu_1\geq\nu_2\geq\nu_3\geq\nu_4\geq0$ are the eigenvalues of the matrix
\begin{equation}
\label{nn60}
\xi(\rho) = \rho(S^y\otimes S^y) \rho^*(S^y\otimes S^y).
\end{equation}
Here, $\rho^*$ is obtained from $\rho$ by representing the latter in the energy basis and then taking the elementwise complex conjugate, and $S^y$ is the Pauli matrix
$
S^y =
\left[
\begin{array}{cc}
0 & -\i \\
\i & 0
\end{array}
\right]$.
The concurrence is related in a monotone way to the {\it entanglement of formation}, and \fer{60} takes values in the interval $[0,1]$. If $C(\rho)=0$ then
the state $\rho$ is separable, meaning that $\rho$ can be written
as a mixture of pure product states. If $C(\rho)=1$ we call $\rho$ maximally entangled.

Let $\rho_0$ be an initial state of $S$. The smallest number $t_0\geq 0$ s.t. $C(\rho_t)=0$ for all $t\geq t_0$ is called the {\em disentanglement time} (also `entanglement sudden death time', \cite{yu1,HZ}). If $C(\rho_t)>0$ for all $t\geq 0$ then we set $t_0=\infty$. The disentanglement time depends on the initial state.
\medskip
Consider the family of pure initial states of $\s$ given by
$$
\rho_0=|\psi\rangle\langle\psi|, \mbox{\qquad with \qquad} \psi = \frac{a_1}{\sqrt{|a_1|^2+|a_2|^2}}\,  |++\rangle \ + \frac{a_2}{\sqrt{|a_1|^2+|a_2|^2}}\,|--\rangle,
$$
where $a_1, a_2\in\cx$ are arbitrary (not both zero). The initial concurrence is
$$
C(\rho_0) =  2\frac{|\Re \,a_1 a_2^*|}{|a_1|^2+|a_2|^2},
$$
which covers all values between zero (e.g. $a_1=0$) to one (e.g. $a_1=a_2\in\rx$). According to \fer{42}, the
density matrix of $\s$ at time $t\geq 0$ is given by
\begin{equation}
\rho_t =
\left[
\begin{array}{cccc}
p_1 & 0 & 0 & \alpha\\
0 & p_2 & 0 & 0 \\
0 & 0 & p_3 & 0 \\
\alpha^* & 0 & 0 & p_4
\end{array}
\right] +O(\varkappa^2),
\label{m3}
\end{equation}
with remainder uniform in $t$, and where $p_j=p_j(t)$ and
$\alpha=\alpha(t)$ are given by the main term on the r.h.s. of
\fer{42}. The initial conditions are $p_1(0) = \frac{|a_1|^2}{|a_1|^2+|a_2|^2}$, $p_2(0)=p_3(0)=0$, $p_4(0) = \frac{|a_2|^2}{|a_1|^2+|a_2|^2}$, and $\alpha(0) = \frac{a_1^* a_2}{|a_1|^2+|a_2|^2}$. We set
\begin{equation}
p:= p_1(0)\in [0,1]
 \label{p}
\end{equation}
and note that
$p_4(0) =1-p$ and $|\alpha(0)| = \sqrt{p(1-p)}$. In terms of $p$, the initial concurrence is
$C(\rho_0)=2\sqrt{p(1-p)}$.
Let us set
\begin{equation}
\delta_2 := (\lambda^2_1+\mu^2_1)\sigma_g(B_1), \qquad \delta_3 := (\lambda^2_2+\mu^2_2)\sigma_g(B_2),
\label{delta2}
\end{equation}
\begin{equation}
\delta_5 := \delta_2+\delta_3 +\left[(\kappa_1+\kappa_2)^2 +\nu_1^2+\nu_2^2\right]\sigma_f(0).
\label{deltas}
\end{equation}
\begin{equation}
\delta_+ := \max\{\delta_2,\delta_3\},\qquad \delta_-:=\min\{\delta_2,\delta_3\}.
\label{deltapm}
\end{equation}
An analysis of the concurrence of \fer{m3}, where the $p_j(t)$ and $\alpha(t)$ evolve according to \fer{42} yields the following bounds on disentanglement time.

\medskip
{\bf Result on disentanglement time.\ }
{\it  Take  $p\neq 0,1$ and suppose that $\delta_2, \delta_3>0$. There is a constant $\varkappa_0>0$
(independent of\ $p$) such that we have:

\medskip
{\bf A.\ }  {\em (Upper bound.)}   There is a constant $C_A>0$ (independent of\
$p,\varkappa$) s.t. $C(\rho_t)=0$ for all $t\geq t_A$, where
\begin{equation}
t_A:= \max\left\{ \frac{1}{\delta_5}\ln\left[C_A
\frac{\sqrt{p(1-p)}}{\varkappa^2}\right],
\frac{1}{\delta_2+\delta_3} \ln\left[C_A
\frac{p(1-p)}{\varkappa^2}\right],
\frac{C_A}{\delta_2+\delta_3}\right\}. \label{m17}
\end{equation}
{\bf B.\ } {\em (Lower bound.)} There is a constant $C_B>0$ (independent of $p$, $\varkappa$) s.t. $C(\rho_t)>0$ for all $t\leq t_B$, where
\begin{equation}
 t_B:=\min\left\{ \frac{1}{\delta_2+\delta_3}\ln[1+ C_B p(1-p)], \frac{1}{\delta_+}\ln\left[1+C_B\varkappa^2 \right], \frac{C_B}{\delta_5-\delta_-/2}
 \right\}. \label{m18}
\end{equation}
}

Bounds \fer{m17} and \fer{m18} are obtained by a detailed analysis of \fer{60}, with $\rho$ replaced by $\rho_t$, \fer{m3}. This analysis is quite straightforward but rather lengthy. Details are presented in \cite{mm}.

\medskip
\noindent
{\bf Discussion.\ } 1. The result gives disentanglement bounds for the true dynamics of the qubits for interactions which are not integrable.

2. The disentanglement time is {\it finite}. This follows from $\delta_2, \delta_3>0$ (which in turn implies that the total system approaches equilibrium as $t\rightarrow\infty$). If the system does not thermalize then it can happen that entanglement stays nonzero for all times (it may decay or even stay constant) \cite{yu1, PR}.

3. The rates $\delta$ are of order $\varkappa^2$. Both $t_A$ and $t_B$ increase with decreasing coupling strength.

4. Bounds \fer{m17} and \fer{m18} are not optimal. The disentanglement time bound \fer{m17} depends on both kinds of couplings. The contribution of each interaction decreases $t_A$ (the bigger the noise the quicker entanglement dies). The bound on entanglement survival time \fer{m18} does not depend on the energy-conserving couplings.

\section{Entanglement creation}
\label{sectentcre}
Consider an initial condition $\rho_\s\otimes\rho_{\r_1}\otimes
\rho_{\r_2}\otimes\rho_{\r_3}$, where $\rho_\s$ is the initial state
of the two qubits, and where the reservoir initial states are
thermal, at fixed temperature $T=1/\beta>0$.

Suppose that the qubits
are not coupled to the collective reservoir $\r_3$, but only to the
local ones, via energy conserving and exchange interactions (local
dynamics). It is not difficult to see that then, if $\rho_\s$ has
zero concurrence, its concurrence will remain zero for all times.
This is so since the dynamics factorizes into parts for $\s_1+\r_1$
and $\s_2+\r_2$, and acting upon an unentangled initial state does
not change entanglement. In contrast, for certain {\it entangled}
initial states $\rho_\s$, one observes death and revival of
entanglement \cite{BLC}: the initial concurrence of the qubits
decreases to zero and may stay zero for a certain while, but it then
grows again to a maximum (lower than the initial concurrence) and
decreasing to zero again, and so on. The interpretation is that
concurrence is shifted from the qubits into the (initially
unentangled) reservoirs, and if the latter are not Markovian,
concurrence is shifted back to the qubits (with some loss).

Suppose
now that the two qubits are coupled only to the collective
reservoir, and not to the local ones. Braun \cite{Braun} has
considered the explicitly solvable model (energy-conserving
interaction), as presented in Section \ref{sectcomp} with
$\kappa=1$, $\nu=0$.\footnote{In fact, Brown uses this model and
sets the Hamiltonian of the qubits equal to zero. This has no
influence on the evolution of concurrence, since the free dynamics
of the qubits can be factored out of the total dynamics
(energy-conserving interaction), and a dynamics of $\s_1$ and $\s_2$
which is a prouct does not change the concurrence.} Using the exact
solution \fer{exact}, Braun calculates the smallest eigenvalue of
the partial transpose of the density matrix of the two qubits, with
$S$ and $\Gamma$ considered as non-negative parameters. For the
initial product state where qubits 1 and 2 are in the states
$\frac{1}{\sqrt 2}(|+\rangle-|-\rangle)$ and $\frac{1}{\sqrt
2}(|+\rangle+|-\rangle)$ respectively, i.e.,
\begin{equation}
\rho_\s = \frac{1}{4}
\left[
\begin{array}{cccc}
1 & 1 & -1 & -1\\
1 & 1 & -1 & -1 \\
-1 & -1 & 1 & 1\\
-1 & -1 & 1 & 1
\end{array}
\right],
\label{instate}
\end{equation}
it is shown that for small values of $\Gamma$ (less than 2,
roughly), the negativity of the smallest eigenvalue of the partial
transpose oscillates between zero and -0.5 for $S$ increasing from
zero. As $\Gamma$ takes values larger than about 3, the smallest
eigenvalue is zero (regardless of the value of $S$). According to
the Peres-Horodecki criterion \cite{peres, horodecki}, the qubits
are entangled exactly when the smallest eigenvalue is strictly below
zero. Therefore, taking into account \fer{S} and \fer{Gamma},
Braun's work \cite{Braun} shows that for small times ($\Gamma$
small) the collective environment (with energy-conserving
interaction) induces first creation, then death and revival of
entanglement in the initially unentangled state \fer{instate}, and
that for large times ($\Gamma$ large), entanglement disappears.

\medskip
{\bf Resonance approximation.\ } The main term of the r.h.s. of \fer{42} can be calculated explicitly, and we give in Appendix A the concrete expressions. How does concurrence evolve under this approximate evolution of the density matrix?

(1) {\it Purely energy-exchange coupling.}\
In this situation we have $\kappa=\nu=0$. The explicit expressions (Appendix A)
show that the density matrix elements $[\rho_t]_{mn}$ in the resonance
approximation depend on $\lambda$ (collective) and $\mu$ (local) through
the symmetric combination $\lambda^2+\mu^2$ only. It follows that
the dominant dynamics \fer{42} (the true dynamics modulo an error
term $O(\varkappa^2)$ homogeneously in $t\geq 0$) is {\it the same}
if we take purely collective dynamics ($\mu=0)$ or purely local
dynamics ($\lambda =0$).
{\it In particular, creation of entanglement under purely collective and
purely local energy-exchange dynamics is
{\em the same}, modulo $O(\varkappa^2)$}.
For instance, for the initial state \fer{instate},
collective energy-exchange couplings can create entanglement of at
most $O(\varkappa^2)$, since local energy-exchange
couplings do not create any entanglement in this initial state.

(2) {\it Purely energy-conserving coupling.}\  In this situation we have $\lambda=\mu=0$. The evolution of the density matrix elements is not symmetric as a function of the coupling constants $\kappa$ (collective) and $\nu$ (local). One may be tempted to conjecture that concurrence is independent of the local coupling parameter $\nu$, since it is so in absence of collective coupling ($\kappa=0$).
However, for $\kappa\neq 0$, concurrence {\it depends} on $\nu$ (see numerical results below). We can understand this as follows. Even if the initial state is unentangled, the collective coupling creates quickly a little bit of entanglement and therefore the local environment does not see a product state any more, and starts processes of creation, death and revival of entanglement.

(3) {\it Full coupling.}\ In this case all of $\kappa, \lambda, \mu, \nu$ do not vanish. Matrix elements evolve as complicated functions of these parameters, showing that the effects of different interactions are correlated.

\section{Numerical Results}
\label{numres}

In the following, we ask whether the resonance approximation is
sufficient to detect creation of entanglement. To this end, we take
the initial condition \fer{instate} (zero concurrence) and study
numerically its evolution under the approximate resonance evolution
(Appendices A, B), and calculate concurrence as a function of time.
Let us first consider the case of purely energy conserving
collective interaction, namely $\lambda=\mu=\nu=0$ and only $\kappa
\ne 0$.
\begin{figure}[!ht]
\begin{center}
\epsfxsize=85mm
\epsfbox{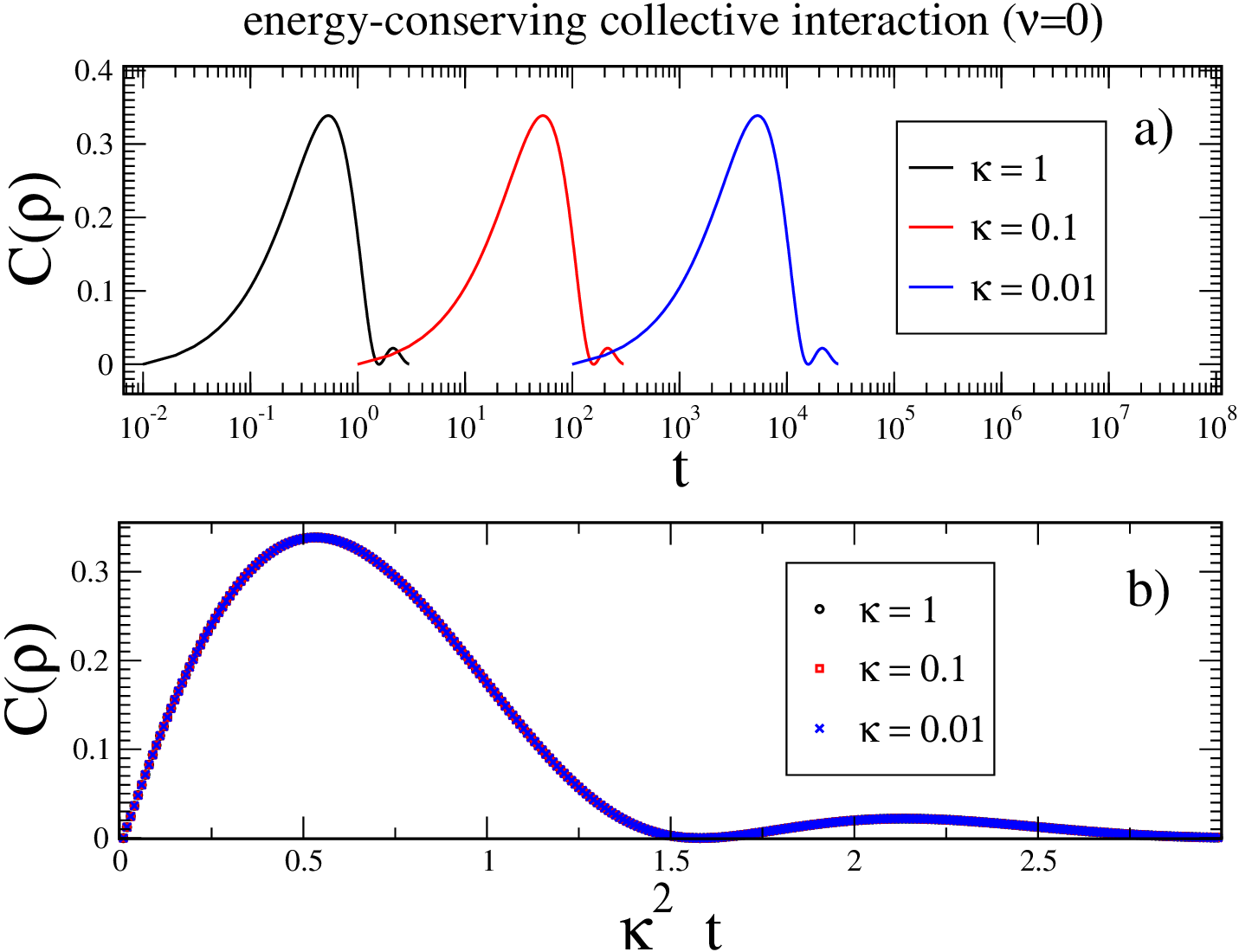}
\end{center}
\caption{{\small Energy conserving collective interaction $\lambda=\mu=\nu=0$.
a) Concurrence a function of time for different $\kappa$ values as indicated in the legend.
b) The same as a) but in the renormalized time $\kappa^2 t $.
}}
\label{f1}
\end{figure}
Our simulations (Figure \ref{f1}a) show that,
a concurrence of value approximately 0.3 is created, independently of the value of
$\kappa$ (ranging from 0.01 to 1). It is clear from the graphs that the effect of
varying $\kappa$ consists only in a time shift. This shift of time is particularly accurate, as can be seen in Fig.~\ref{f1}b, where the three curves drawn in a)
collapse to a single curve under the time rescaling $t \to \kappa^2 t $.
In particular, the maximum concurrence is taken at times $t_0\approx 0.5\kappa^{-2}$. We also point
out that the revived concurrence has very small amplitude (approximately 15 times smaller than
the maximum concurrence)
and takes its maximum at $t_1\approx 2.1\kappa^{-2}$. Even though the amplitude of the revived concurrence is small as compared to $\kappa^2$, the graphs show that it is {\it independent} of $\kappa$, and hence our resonance dynamics does reveal concurrence revival.
\begin{figure}[!ht]
\begin{center}
\epsfxsize=85mm
\epsfbox{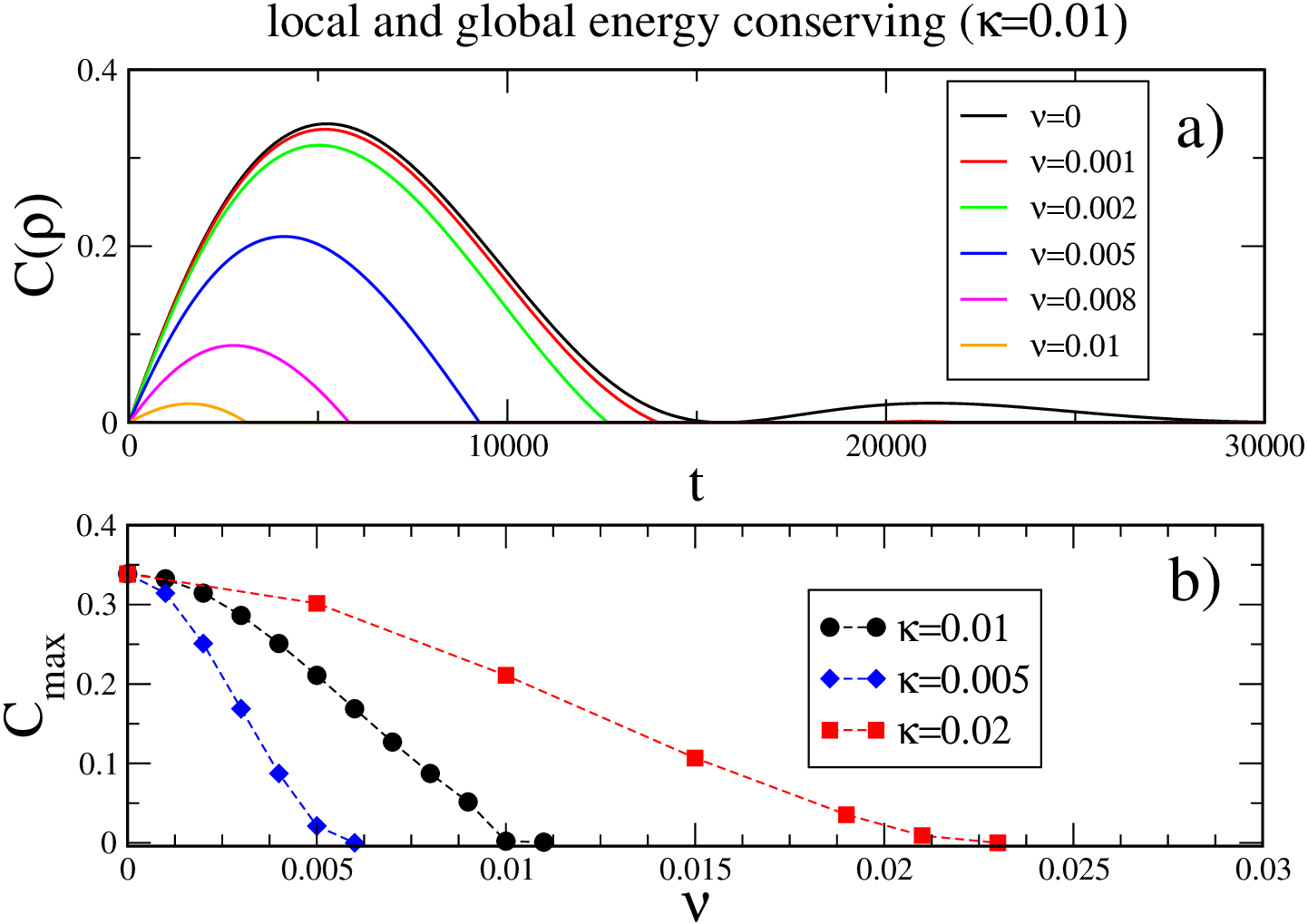}
\end{center}
\caption{{\small Energy conserving collective and local interaction $\lambda=\mu=0$.
a) Concurrence a function of time for fixed collective interaction  $\kappa = 0.01$
and different local interaction $\nu$  as indicated in the legend.
b) Variation of the maximum of concurrence as a function of the local
interaction strength $\nu$ for different collective interaction strengths $\kappa$ as indicated in the legend.
}}
\label{f2}
\end{figure}

When switching on the local energy conserving coupling, $\nu\neq 0$, we see
in Fig.~\ref{f2}a,
that the maximum of concurrence decreases with increasing $\nu$.
Therefore, the effect of a local coupling is to reduce the entanglement. It is also interesting
to study the dependence of the maximal value of the concurrence, $C_{\rm max}$, as a function of the energy-conserving interaction parameters.  This is done in Fig.~\ref{f2}b, where $C_{\rm max}$ is plotted as a function of the local interaction $\nu$, for different fixed collective
couplings $\kappa$. The graphs show that as the local coupling $\nu$ is increased to the value of the collective coupling $\kappa$, $C_{\rm max}$ becomes zero. This means that if the local coupling exceeds the collective one, then there is no creation of concurrence. We may interpret this as a competition between the concurrence-reducing tendency of the local coupling (apart from very small revival effects) and the concurrence-creating tendency of the collective coupling (for not too long times). If the local coupling exceeds the collective one, then concurrence is prevented from building up.
\begin{figure}[!ht]
\begin{center}
\epsfxsize=85mm
\epsfbox{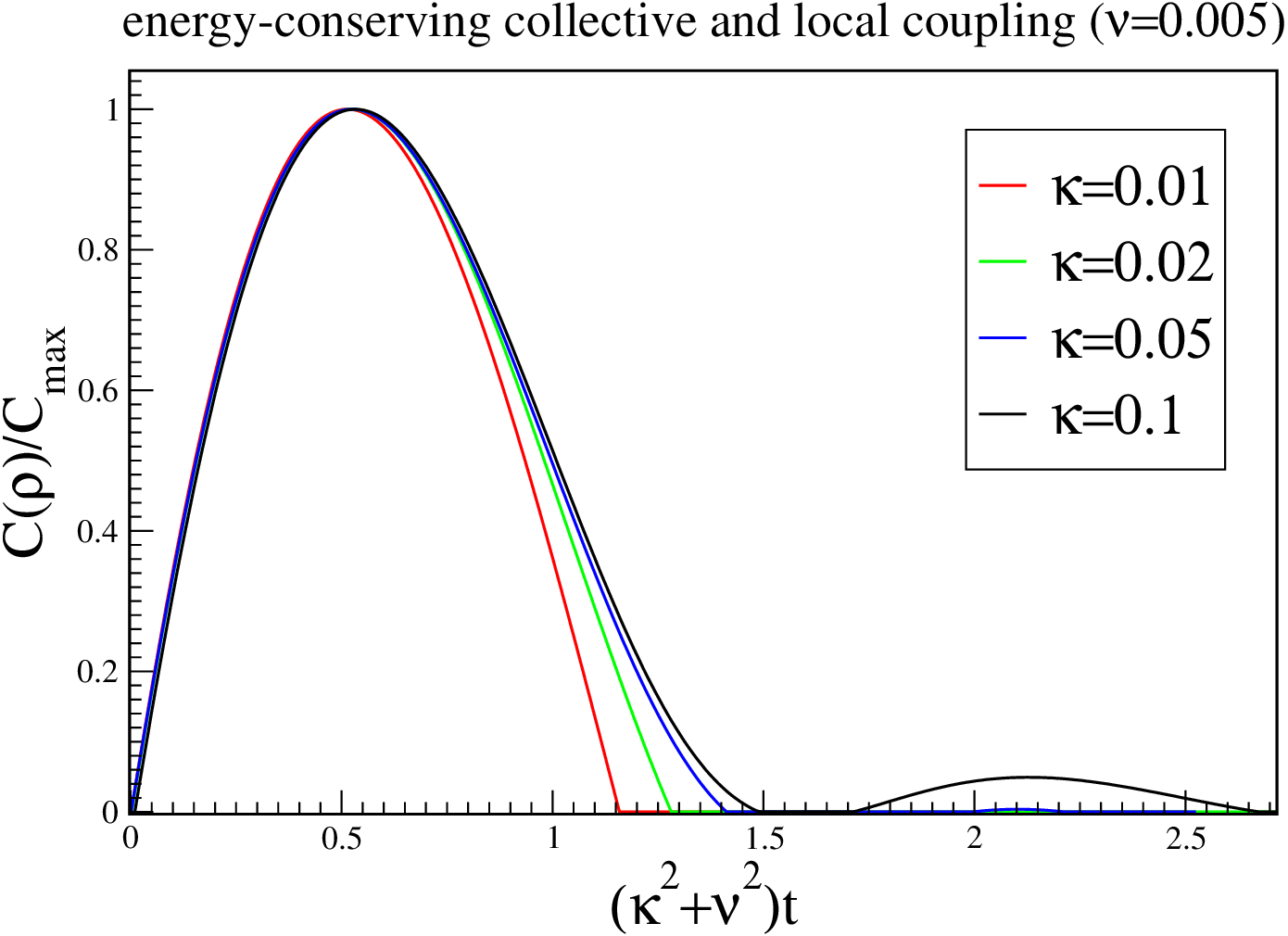}
\end{center}
\caption{{\small Energy conserving collective and local interaction $\lambda=\mu=0$.
Rescaled concurrence $C(\rho)/C_{max}$ as function of time for fixed local interaction  $\nu = 0.005$
and different collective interaction $\kappa > \nu $  (as indicated in the legend)  as a function
of the rescaled time $(\kappa^2 + \nu^2) t $.
}}
\label{f3}
\end{figure}

\begin{figure}[!ht]
\begin{center}
\epsfxsize=85mm
\epsfbox{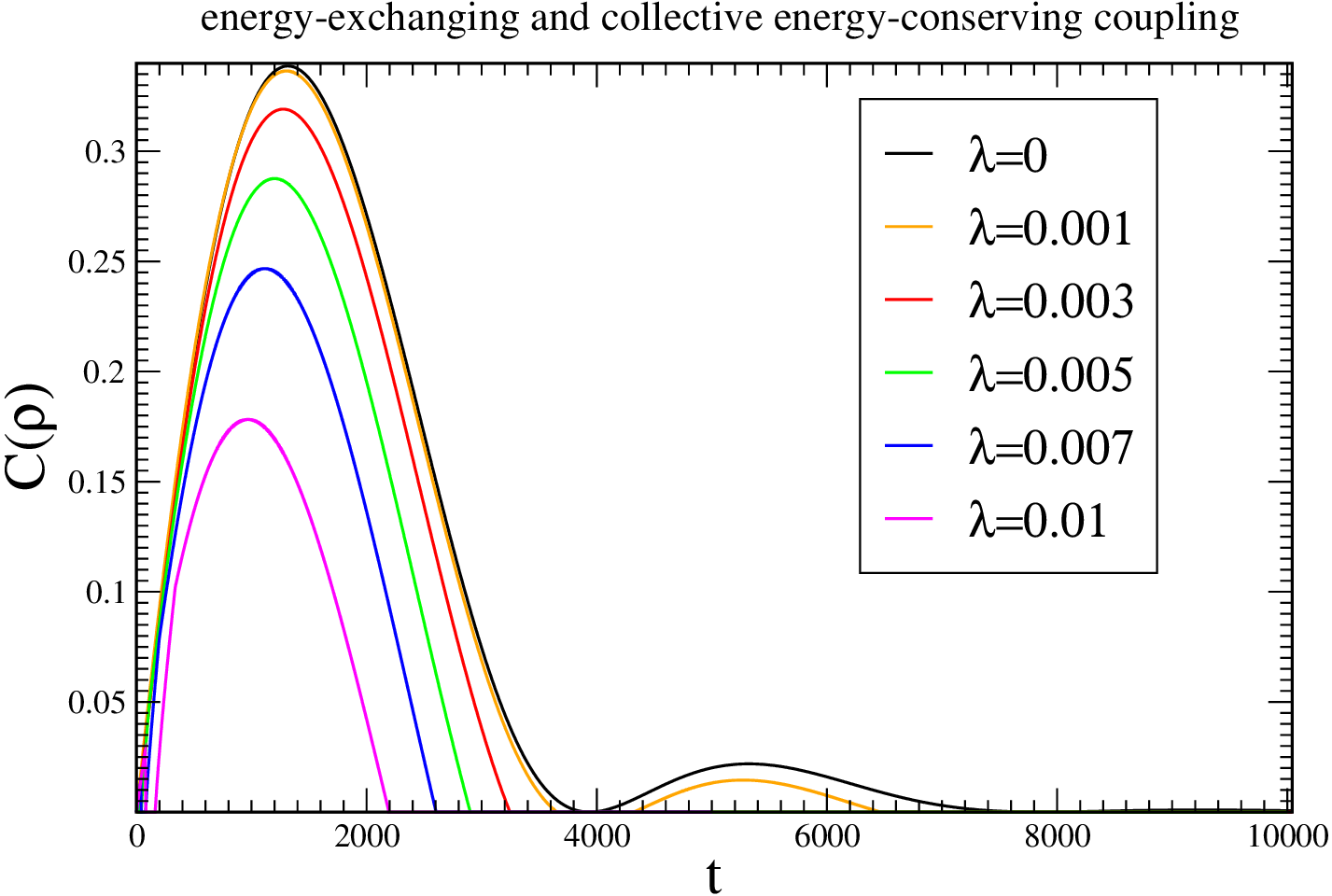}
\end{center}
\caption{{\small Energy-exchanging collective and local interactions $\lambda=\mu\ne 0$.
Concurrence $C(\rho)$ as function of time for fixed energy-conserving
collective interactions  $\kappa=0.02$, $\nu=0$
and different energy-exchanging couplings $\lambda$ as indicated in the legend.
Here we used $B_1 = 1$, $B_2 = 1.25$, and $\beta = 1$.
}}
\label{f4}
\end{figure}

Looking at Fig.~\ref{f2}, it is clear that the effect of the local coupling is not only
to decrease concurrence but also to induce a shift of time, similarly to the effect
of the collective coupling $\kappa$. Indeed, taking as a variable the rescaled
concurrence $C(\rho)/C_{max}$, one can see that the approximate scaling $(\kappa^2+\nu^2)t$ is at work, see Fig.~\ref{f3}.
{\it We conclude that both local and collective energy conserving interactions produce a cooperative time shift of the entanglement creation, but only the local interaction can destroy entanglement creation. There is no entanglement creation for $\nu > \kappa$.}

\medskip
Let us now consider an additional energy exchange coupling $\lambda, \mu \ne 0$.
Since these parameters appear in the resonance dynamics only in
the combination $\lambda^2+\mu^2$, see Appendix A, we set without
loosing generality $\lambda = \mu$. We plot in Fig.~\ref{f4} the time evolution of
the concurrence, at fixed energy-conserving couplings $\kappa=0.02$ and $\nu=0$,
for different values of the energy exchange coupling $\lambda$.
In this case we have chosen $B_1=1$ which
corresponds to $\omega_0=\omega_1/2$, where $\omega_1$ is a
transition frequency of the first qubit. We also used the
conditions: $\sigma_g(B_1)=r_g(B_1)=1$, which lead to the
renormalization of the interaction constants. The relations between
$\sigma_g(B_2)$ and $\sigma_g(B_1)$, and $r_g(B_2)$ and $r_g(B_1)$
are discussed in Appendix B.

\begin{figure}[!ht]
\begin{center}
\epsfxsize=85mm
\epsfbox{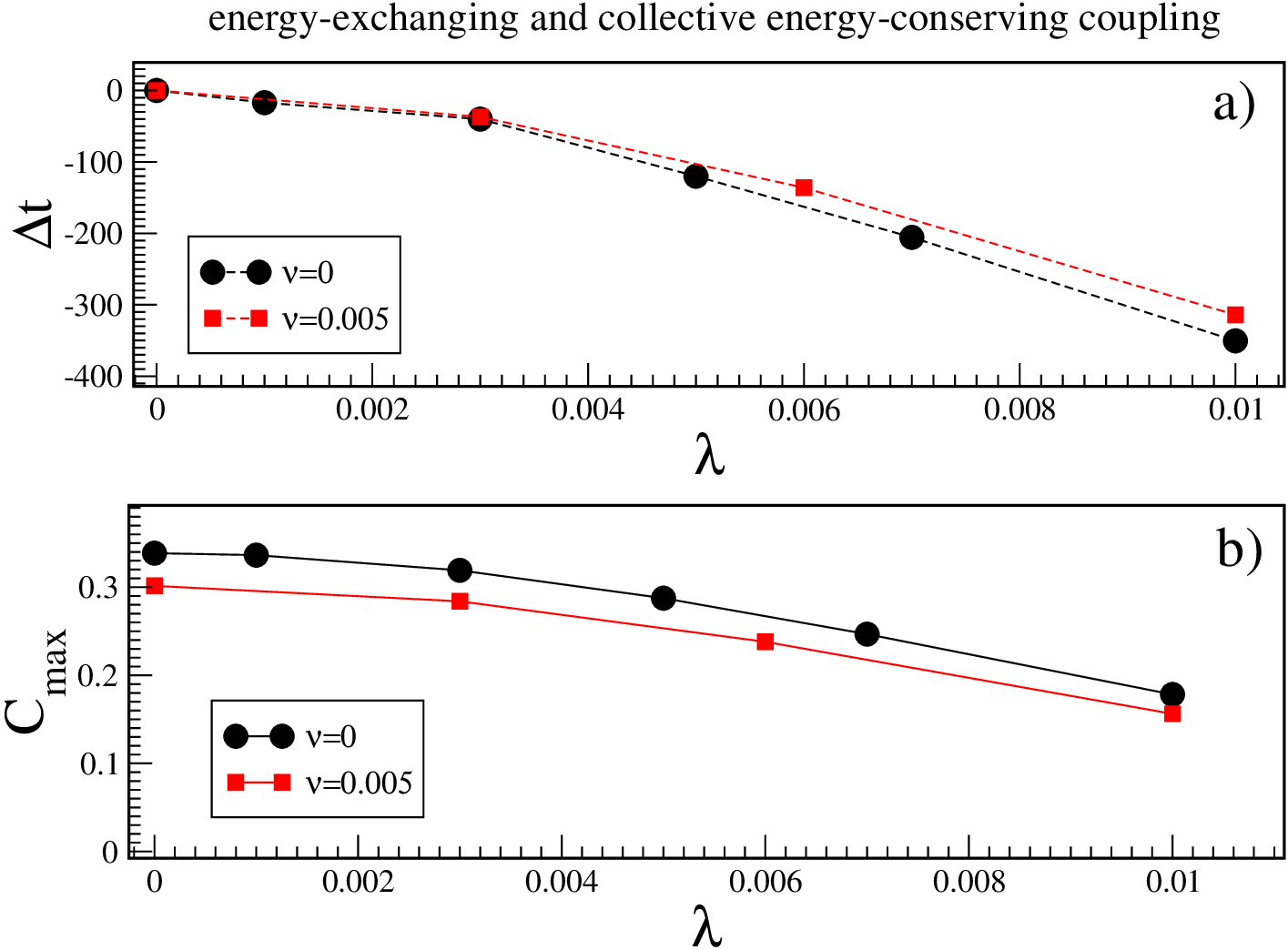}
\end{center}
\caption{{\small Energy-exchanging collective and local interaction $\lambda=\mu\ne 0$.
a) Time shift induced by energy-exchanging coupling, for the same
energy conserving collective coupling $\kappa=0.02$ and different local couplings $\nu$
as indicated in the legend. b) Decay of the maximal concurrence as a function of $\lambda$,
for the same cases as (a). Magnetic fields and temperature is the same as in Fig.~\ref{f4}.
}}
\label{f5}
\end{figure}

Figure \ref{f4} shows that the effect of the energy exchanging coupling is to shift
slightly the time where concurrence is maximal and, at the same time, to decrease the amplitude of concurrence for each fixed time. This feature is analogous to the effect of local energy-conserving interactions, as discussed above.
Unfortunately, it is quite difficult in this case to extract the
threshold values of $\lambda$ at which the creation of concurrence is prevented
for all times. The difficulty comes from the fact that for larger values of $\lambda$, the
concurrence is very small and the negative eigenvalues on order $O(\varkappa^2)$ do not allow
a reliable calculation.
 This picture does not change much if a local energy-conserving interaction $\nu < \kappa$ is added. In Fig.~\ref{f5}, we show respectively,
the time shift of the maximal concurrence
$\Delta t = t_{max}(\lambda) - t_{max} (\lambda=0)$ as a function of the
energy-exchanging coupling $\lambda$ (a)
and the behavior of the maximal concurrence as a function of the same parameter
$\lambda$   for two different values of the local
coupling $\nu$. Is appears evident that the role played by the energy-exchange
coupling is very similar to that played by the local energy-conserving one.

\begin{figure}[!ht]
\begin{center}
\epsfxsize=85mm
\epsfbox{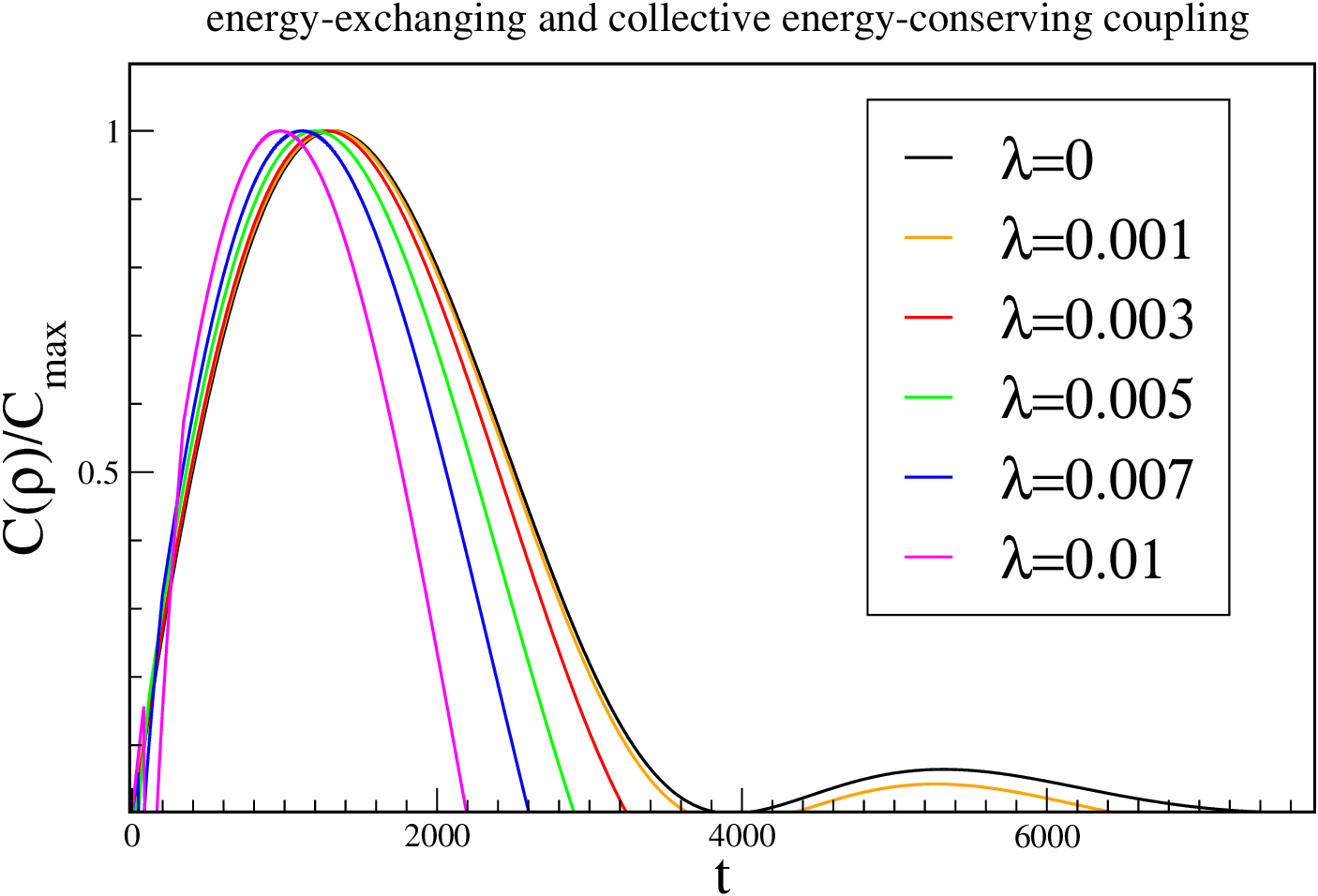}
\end{center}
\caption{{\small Energy-exchanging collective and local interaction $\lambda=\mu\ne 0$.
Rescaled concurrence $C(\rho)/C_{max}$ {\it vs  } time $t$, for different
$\lambda $ values. Here,$\kappa = 0.02$ and $\nu=0$.
Magnetic fields and temperature is the same as in Fig.~\ref{f4}.
}}
\label{f6}
\end{figure}

Let us comment about concurrence revival. The effect of a collective
energy-conserving coupling consists of creating entanglement, destroying it
and creating it again but with a smaller amplitude. Generally speaking,
an energy-exchanging coupling, if extremely small, does not change this picture.
Nevertheless, it is important to stress that the damping effect the energy-exchange
coupling has on the concurrence amplitude is stronger on the revived
concurrence than on the initially created one. This is shown in
Fig.~\ref{f6}, where the renormalized concurrence $C(\rho)/C_{max}$ is plotted
for different $\lambda $ values.
For these parameter values, only a very small
coupling $\lambda \leq 0.001$ will allow revival of concurrence.

In the calculation of concurrence, the square roots of the eigenvalues of the matrix
$\xi(\rho)$ (\ref{nn60}) should be taken. As explained before, the non positivity, to order $O(\varkappa^2)$
of the density matrix $\rho$ reflects on the non positivity of the eigenvalues of the matrix
$\xi(\rho)$.
When this happens ($\nu_i < 0 $) we simply put $\nu_i = 0$ in the numerical calculations. This produces
an approximate (order $O(\varkappa^2)$) concurrence which produces spurious effects,
especially for small time, when concurrence is small. These effects are particularly
evident in Fig.~\ref{f6}, for small time, where artificial oscillations occur, instead
of an expected  smooth behavior. In contrast to this behaviour, the revival of entanglement as revealed in Figure 6 varies smoothly in $\lambda$, indicating that this effect is not created due to the approximation.

\section{Conclusion}
We consider a system of two qubits interacting with local and
collective thermal  quantum reservoirs. Each qubit is coupled to its
local reservoir by two channels, an energy-conserving and an
energy-exchange one. The qubits are collectively coupled to a third
reservoir, again through two channels. This is thus a versatile
model, describing local and collective, energy-conserving and
energy-exchange processes.

We present an approximate dynamics which describes the evolution of
the reduced  density matrix for all times $t\geq 0$, modulo an error
term $O(\varkappa^2)$, where $\varkappa$ is the typical coupling
strength between a single qubit and a single reservoir. The error
term is controlled rigorously and for all times. The approximate
dynamics is markovian and shows that different parts of the reduced
density matrix evolve together, but independently from other parts.
This partitioning of the density matrix into {\it clusters} induces
a classification of decoherence times -- the time-scales during
which a given cluster stays populated. We obtain explicitly the
decoherence and relaxation times and show that their leading
expressions (lowest nontrivial order in $\varkappa$) is independent
of the ultraviolet behaviour of the system, and in particular,
independent of any ultraviolet cutoff, artificially needed to make
the models mathematically well defined.

We obtain analytical estimates on entanglement death and
entanglement  survival times for a class of initially entangled
qubit states, evolving under the full, not explicitly solvable
dynamics. We investigate numerically the phenomenon of entanglement
creation and show that the approximate dynamics, even though it is
markovian, {\it does} reveal creation, sudden death and revival of
entanglement. We encounter in the numerical study a disadvantage of
the approximation, namely that it is not positivity preserving,
meaning that for small times, the approximate density matrix has
slightly negative eigenvalues.

The above-mentioned cluster-partitioning of the density matrix is
valid  for general $N$-level systems coupled to reservoirs. We think
this clustering will play a useful and important role in the
analysis of quantum algorithms. Indeed, it allows one to separate
``significant" from ``insignificant" quantum effects, especially
when dealing with large quantum registers for performing quantum
algorithms. Depending on the algorithm, fast decay of some blocks of
the reduced density matrix elements can still be tolerable for
performing the algorithm with high fidelity.

We point out a further possible application of our method to
novel quantum measuring technologies based on superconducting
qubits. Using two superconducting qubits as measuring devices
together with the scheme considered in this paper will allow one to
extract not only the special density of noise, but also possible
quantum correlations imposed by the environment. Modern methods of
quantum state tomography will allow to resolve these issues.

\appendix
\section{Dynamics in resonance approximation}
We take $0<B_1<B_2$, $B_2/B_1\neq 2$, and $\varkappa^2<\!\!<
\min\{2B_1,2(B_2-B_1), 2|B_2-2B_1|\}$. These conditions guarantee
that the resonances do not overlap, see also \cite{MSB1}. In the
sequel, $\doteq$ means equality modulo an error term
$O(\varkappa^2)$ which is homogeneous in $t\geq 0$. The main
contribution of the dynamics in (\ref{42}) is given as follows.
\begin{eqnarray}
{}[\rho_t]_{11} &\doteq& \frac{1}{Z}\frac{1}{\sqrt{e_1e_2}}\Big\{ (1+\e^{-t\delta_2}e_2 +\e^{-t\delta_3}e_1 +\e^{-t\delta_4}e_1e_2)[\rho_0]_{11}\nonumber\\
&& \qquad+ (1-\e^{-t\delta_2} +\e^{-t\delta_3}e_1 -\e^{-t\delta_4}e_1)[\rho_0]_{22}\nonumber\\
&& \qquad+ (1+\e^{-t\delta_2}e_2 -\e^{-t\delta_3}e_1 -\e^{-t\delta_4}e_2)[\rho_0]_{33}\nonumber\\
&& \qquad+ (1-\e^{-t\delta_2} -\e^{-t\delta_3} -\e^{-t\delta_4})[\rho_0]_{44}\Big\} \label{el11}\\
{}[\rho_t]_{22} &\doteq& \frac{1}{Z}\sqrt{\frac{e_2}{e_1}}\Big\{ (1-\e^{-t\delta_2}+\e^{-t\delta_3}e_1 -\e^{-t\delta_4}e_1)[\rho_0]_{11}\nonumber\\
&& \qquad+ (1+\e^{-t\delta_2}e_2^{-1} +\e^{-t\delta_3}e_1 +\e^{-t\delta_4}e_1e_2^{-1})[\rho_0]_{22}\nonumber\\
&& \qquad+ (1-\e^{-t\delta_2} -\e^{-t\delta_3} +\e^{-t\delta_4})[\rho_0]_{33}\nonumber\\
&& \qquad+ (1+\e^{-t\delta_2}e_2^{-1} -\e^{-t\delta_3} -\e^{-t\delta_4}e_2^{-1})[\rho_0]_{44}\Big\} \label{el22}\\
{}[\rho_t]_{33} &\doteq& \frac{1}{Z}\sqrt{\frac{e_1}{e_2}}\Big\{ (1+\e^{-t\delta_2}e_2-\e^{-t\delta_3} -\e^{-t\delta_4}e_2)[\rho_0]_{11}\nonumber\\
&& \qquad+ (1-\e^{-t\delta_2}-\e^{-t\delta_3}+\e^{-t\delta_4})[\rho_0]_{22}\nonumber\\
&& \qquad+ (1+\e^{-t\delta_2}e_2 +\e^{-t\delta_3}e_1^{-1} -\e^{-t\delta_4}e_2e^{-1}_1)[\rho_0]_{33}\nonumber\\
&& \qquad+ (1-\e^{-t\delta_2} +\e^{-t\delta_3}e_1^{-1} -\e^{-t\delta_4}e_1^{-1})[\rho_0]_{44}\Big\} \label{el33}\\
{}[\rho_t]_{44} &\doteq& \frac{1}{Z}\sqrt{e_1 e_2}\Big\{ (1-\e^{-t\delta_2}-\e^{-t\delta_3} +\e^{-t\delta_4})[\rho_0]_{11}\nonumber\\
&& \qquad+ (1+\e^{-t\delta_2}e_2^{-1}-\e^{-t\delta_3}-\e^{-t\delta_4}e_2^{-1})[\rho_0]_{22}\nonumber\\
&& \qquad+ (1-\e^{-t\delta_2} +\e^{-t\delta_3}e_1^{-1} -\e^{-t\delta_4}e^{-1}_1)[\rho_0]_{33}\nonumber\\
&& \qquad+ (1+\e^{-t\delta_2}e_2^{-1} +\e^{-t\delta_3}e_1^{-1} +\e^{-t\delta_4}e_1^{-1}e_2^{-1})[\rho_0]_{44}\Big\}. \label{el44}
\end{eqnarray}
Here,
\begin{eqnarray}
Z&=&{\rm Tr}\e^{-\beta H_\s},\\
e_j&=&\e^{2\beta B_j}\\
\delta_2 &=& (\lambda^2 +\mu^2)\sigma_g(B_2)\\
\delta_3 &=& (\lambda^2+\mu^2)\sigma_g(B_1)\\
\delta_4 &=&\delta_2+\delta_3.
\label{new99}
\end{eqnarray}
Of course, the populations do not depend on any energy-conserving parameter.
The cluster of matrix elements $\{(3,1), (4,2)\}$ evolves as
\begin{eqnarray}
{}[\rho_t]_{42} &\doteq& \e^{\i t\varepsilon_{2B_1}^{(1)}} \frac{e_2y_+}{1+e_2(y_+)^2} \Big\{ [\rho_0]_{31} + y_+[\rho_0]_{42}\Big\}\nonumber\\
&& + \e^{\i t\varepsilon_{2B_1}^{(2)}} \frac{e_2 y_-}{1+e_2(y_-)^2} \Big\{ [\rho_0]_{31} +y_-[\rho_0]_{42}\Big\} \label{el42},\\
{}[\rho_t]_{31} &\doteq& \e^{\i t\varepsilon_{2B_1}^{(1)}} \frac{1}{1+e_2(y_+)^2} \Big\{ [\rho_0]_{31} + y_+[\rho_0]_{42}\Big\}\nonumber\\
&& + \e^{\i t\varepsilon_{2B_1}^{(2)}} \frac{1}{1+e_2(y_-)^2} \Big\{ [\rho_0]_{31} +y_-{}[\rho_0]_{42}\Big\}. \label{el31}
\end{eqnarray}
Here,
\begin{equation}
\varepsilon_{2B_1}^{(k)} = A+\frac 12 B(1+e_2) - (-1)^k\frac 12 \Big[ B^2(1+e_2)^2 +4C(B(e_2-1)+C)\Big]^{1/2},
\label{100}
\end{equation}
where
\begin{eqnarray}
A&=& \i(\lambda^2+\mu^2)\frac12\sigma_g(B_1) +\i (\kappa^2+\nu^2)\sigma_f(0) -(\lambda^2+\mu^2)r_g(B_1)\\
B&=& \i(\lambda^2+\mu^2)\sigma_g^-(B_2)\\
C&=&-2\kappa^2 r_f\\
y_{\pm} &=& 1+\frac{A+C-\varepsilon_{2B_1}^{(k)}}{e_2 B}\qquad (k=1 \ {\rm for} \ y_+, \ k=2 \
{\rm for} \ y_-).
\end{eqnarray}
and
\begin{equation}
\begin{array}{lll}
\sigma_g(x) &=& \displaystyle
4\pi  x^2 \coth(\beta x) \int_{S^2} |g(2x,
\Sigma)|^2\d\Sigma\\
&&\\
\sigma_g^- (x) &=&  \displaystyle  2 \pi
 x^2\frac{\e^{\beta x}}{\sinh(\beta x)} \int_{S^2} |g(2x) ,\Sigma)|^2\d\Sigma \\
&&\\
r_g(x) &=&\frac12\ {\rm P.V.}\int_{\rx\times S^2} u^2|g(|u|,\Sigma)|^2\coth(\beta|u|/2)
 \frac{1}{ u-2x} \d u \d\Sigma\\
&&\\
r_f &=& {\rm P.V.} \displaystyle \int_{\rx^3}\frac{|f|^2}{|k|}\d^3k\\
&&\\
\sigma_f(0) &=& \displaystyle 4\pi
 \lim_{x\downarrow 0} x^2 \coth(\beta x) \int_{S^2} |f(2x),\Sigma)|^2\d\Sigma.
\label{unk}
\end{array}
\end{equation}
The cluster of matrix elements $\{(2,1), (4,3)\}$ evolves as
\begin{eqnarray}
{}[\rho_t]_{21} &\doteq& \e^{\i t\varepsilon_{2B_2}^{(1)}} \frac{1}{1+e_1(y'_+)^2} \Big\{ [\rho_0]_{21} + y'_+[\rho_0]_{43}\Big\}\nonumber\\
&& + \e^{\i t\varepsilon_{2B_2}^{(2)}} \frac{1}{1+e_1(y'_-)^2} \Big\{ [\rho_0]_{21} +y_-[\rho_0]_{43}\Big\} \label{el21},\\
{}[\rho_t]_{43} &\doteq& \e^{\i t\varepsilon_{2B_2}^{(1)}} \frac{e_1 y'_+}{1+e_1(y'_+)^2} \Big\{ [\rho_0]_{21} + y'_+[\rho_0]_{43}\Big\}\nonumber\\
&& + \e^{\i t\varepsilon_{2B_2}^{(2)}} \frac{e_1 y'_-}{1+e_1(y'_-)^2} \Big\{ [\rho_0]_{21} +y'_-{}[\rho_0]_{43}\Big\}. \label{el43}
\end{eqnarray}
Here, $\varepsilon_{2B_2}^{(k)}$ is the same as
$\varepsilon_{2B_1}^{(k)}$, but with all indexes labeling qubits 1
and 2 interchanged ($e_1\leftrightarrow e_2$, $B_1\leftrightarrow
B_2$ in all coefficients involved in  $\varepsilon_{2B_1}^{(k)}$
above). Also, $y'_\pm$ is obtained from $y_\pm$ by the same switch
of labels. Finally,
\begin{eqnarray}
{}[\rho_t]_{32} &\doteq& \e^{\i t\varepsilon_{2(B_1-B_2)}} [\rho_0]_{32} \label{el32}\\
{}[\rho_t]_{41} &\doteq& \e^{\i t\varepsilon_{2(B_1+B_2)}} [\rho_0]_{41} \label{el41}
\end{eqnarray}
with
\begin{eqnarray*}
\varepsilon_{2(B_1-B_2)} &=& \i (\lambda^2 +\mu^2) [\sigma_g(B_1)+\sigma_g(B_2)] +2\i\nu^2\sigma_f(0)\nonumber\\
&& +(\lambda^2+\mu^2) [r_g(B_1)-r_g(B_2)]\\
\varepsilon_{2(B_1+B_2)} &=& \i (\lambda^2 +\mu^2) [\sigma_g(B_1)+\sigma_g(B_2)] +4\i\kappa^2 \sigma_f(0) + 2\i\nu^2\sigma_f(0)\nonumber\\
&& -(\lambda^2+\mu^2) [r_g(B_1)+r_g(B_2)].
\end{eqnarray*}

\section{Reduction to independent parameters}

The equations above contain four independent coupling constants
$\lambda, \mu, \nu, \kappa$ describing the energy-conserving and the
energy exchanging (local and collective) interaction, and eight
different functions of the form factors $f$ and $g$ : $\sigma_g
(B_i)$, $r_g(B_i)$, $\sigma_g^- (B_i)$, $i=1,2$, $\sigma_f(0)$,
$r_f$ (\ref{unk}).

These functions are not independent. First of all it is easy to see
that the following relation holds:
\begin{equation}
\label{rip1}
\sigma_g^- (x) = \frac{ e^{2\beta x}}{e^{2\beta x}+1} \sigma_g(x),
\end{equation}
moreover, choosing for instance a form factor $g(2x, \Sigma ) \propto \sqrt{2x}$ one has:
\begin{equation}
\frac{\sigma_g (B_2)}{\sigma_g (B_1)} = \left( \frac{B_2}{B_1} \right)^3 \frac{ \coth \beta B_2}
{\coth \beta B_1} .
\label{rip2}
\end{equation}
Integrals in $du$ in Eq.~(\ref{unk}) converge only when adding a
cut-off $u_c$. It is easy to show that, when $u_c \to \infty$ one
has:
\begin{equation}
\lim_{u_c\to \infty} \frac{r_g (B_2)}{r_g (B_1)} = 1,
\label{rip3}
\end{equation}
and we can assume $r_g(B_1) \simeq r_g (B_2)$. So, we end up with
four independent divergent integrals, $\sigma_g(B_1), \  r_g(B_1), \
\sigma_f(0), \ r_f,$ in terms of which we can write explicitly the
decay rates :
\begin{equation}
\begin{array}{lll}
\alpha_1 &=& (\lambda^2+\mu^2) \sigma_g(B_1)\\
&&\\
\alpha_2 &=& (\lambda^2+\mu^2) \sigma_g(B_1)
 \left( \frac{B_2}{B_1} \right)^3 \frac{ \coth \beta B_2}
{\coth \beta B_1}\\
&&\\
\alpha_3 &=& \kappa^2 \sigma_f (0)\\
&&\\
\alpha_4 &=& \nu^2 \sigma_f(0),
\end{array}
\end{equation}
and the Lamb shifts,
\begin{equation}
\begin{array}{lll}
\beta_1 &=& (\lambda^2+\mu^2) r_g(B_1)\\
&&\\
\beta_2 &=& (\lambda^2+\mu^2) r_g(B_2) \simeq \beta_1\\
&&\\
\beta_3 &=& -\kappa^2 r_f .
\end{array}
\end{equation}

Suppose now that both Lamb shifts, and decay constants are
 {\it experimentally measurable}
quantities,  and also assume (due to symmetry) that  $\lambda=\mu$.
Interaction constants can be renormalized in order to give directly
decay constants and Lamb shifts:
\begin{equation}
\begin{array}{lll}
\alpha_1 &=&  2 \tilde{\lambda}^2\\
&&\\
\alpha_2 &=& 2 \tilde{\lambda}^2
 \left( \frac{B_2}{B_1} \right)^3 \frac{ \coth \beta B_2}
{\coth \beta B_1}\\
&&\\
\alpha_3 &=& \tilde{\kappa}^2 \\
&&\\
\alpha_4 &=& \tilde{\nu}^2,\\
\beta_1 &=& 2 \tilde{\lambda}^2\\
&&\\
\beta_2 &=&  \beta_1\\
&&\\
\beta_3 &=& -\tilde{\kappa}^2.
\end{array}
\end{equation}

$\tilde{\lambda}, \ \tilde{\kappa}, \tilde{\nu}$ are the values chosen for simulations.


\begin{thebibliography}{}

\bibitem{yu1}
T. Yu and J. H. Eberly,  Qubit disentanglement and decoherence via dephasing, {\em Phys. Rev. B}, {\bf 68}, 165322-1-9 (2003), Yu, T., Eberly, J.H.: Finite-Time Disentanglement Via Spontaneous Emission. {\em Phys. Rev. Lett.} {\bf 93}, no.14, 140404 (2004); Sudden Death of Entanglement. {\em Sience}, {\bf 323}, 598-601, 30 January 2009; Sudden death of entanglement: Classical noise effects. {\em Optics Communications}, {\bf 264}, 393-397 (2005).

\bibitem{wbps}
J. Wang, H. Batelaan, J. Podany, and A.F Starace, Entanglement evolution in the presence of decoherence, {\em J. Phys. B: At. Mol. Opt. Phys.} {\bf 39},  4343-4353 (2006).

\bibitem{AFOV}
L. Amico, R. Fazio, A. Osterloh, and V. Vedral, Entanglement in many-body systems, {\em Rev. Mod. Phys.}, {\bf 80}, 518- 576 (2008).

\bibitem{FLWDR}
O.J. Far\!\'{\!\!i}as, C.L. Latune, S.P. Walborn, L. Davidovich, and P.H.S. Ribeiro, Determining the dynamics of entanglement,  {\em SCIENCE}, {\bf 324}, 1414-1417 (2009).

\bibitem{MSS}
Y.  Makhlin, G. Sch\"on, and A. Shnirman, Quantum-state engineering with Josephson-junction devices, {\em Rev. Mod. Phys.}, {\bf 73}, 380-400 (2001).

\bibitem{YHCCW}
Y. Yu, S. Han, X. Chu, S.I Chu,  Z. Wang, Coherent temporal oscillations of macroscopic quantum states in a Josephson junction, {\em SCIENCE},  {\bf 296}, 889-892 (2002).

\bibitem{DM}
M.H. Devoret and J.M. Martinis, Implementing qubits with superconducting integrated circuits, {\em Quantum Information Processing}, {\bf 3}, 163-203 (2004).

\bibitem{Setal}
M. Steffen, M. Ansmann, R. McDermott, N. Katz, R.C. Bialczak, E. Lucero, M. Neeley, E.M. Weig, A.N. Cleland, and J.M. Martinis, State tomography of capacitively shunted phase qubits with high fidelity, {\em Phys. Rev. Lett.} {\bf  97}  050502-1-4 (2006).

\bibitem{Ketal}
N. Katz,  M. Neeley, M. Ansmann, R.C. Bialczak,  M. Hofheinz, E. Lucero,  A. O'Connell,  Reversal of the weak measurement of a quantum state in a superconducting phase qubit, {\em Phys. Rev. Lett.}, {\bf  101}, 200401-1-4 (2008).

\bibitem{Cletal}
A.A. Clerk, M.H. Devoret, S.M. Girvin, Florian Marquardt, and R.J. Schoelkopf, Introduction to quantum noise, measurement and amplification, {\em Preprint arXiv:0810.4729v1 [cond-mat]} (2008).


\bibitem{MSB1}
M. Merkli, I.M. Sigal, G.P. Berman:  Decoherence and thermalization. {\em Phys. Rev. Lett.} {\bf  98}  no. 13, 130401, 4 pp (2007);\ Resonance theory of decoherence and thermalization. {\em Ann. Phys.} {\bf 323}, 373-412 (2008);\ Dynamics of collective decoherence and thermalization. {\em Ann. Phys.} {\bf 323}, no. 12, 3091-3112 (2008).


\bibitem{mm}
M. Merkli: Entanglement Evolution, a Resonance Approach. Preprint 2009.

\bibitem{PSE}
G.M. Palma, K.-A. Suominen, A.K. Ekert: Quantum Computers and Dissipation.
{\em Proc. R. Soc. Lond. A} {\bf 452}, 567-584 (1996).


\bibitem{MLSO}
M. Merkli: Level shift operators for open quantum systems. {\em J. Math. Anal. Appl.} {\bf 327}, Issue 1, 376-399 (2007).


\bibitem{BP}
H.-P. Breuer, F. Petruccione: The theory of open quantum systems. Oxford university press 2002.


\bibitem{BLC}
B. Bellomo, R. Lo Franco, G. Compagno: Non-Markovian Effects on the Dynamics of Entanglement. {\em Phys. Rev. Lett.} {\bf 99}, 160502 (2007).



\bibitem{Leggett}
A.J. Leggett, S. Chakravarty, A.T. Dorsey, Matthew P.A. Fisher,  Anupam Garg, W. Zwerger: Dynamics of the dissipative two-state system. {\it
Rev. Mod. Phys.} {\bf 59}, 1–85 (1987).


\bibitem{SMS}
A. Shnirman, Yu. Makhlin, G. Sch\"on: Noise and Decoherence in Quantum Two-Level Systems. {\it Physica Scripta} {\bf T102}, 147-154 (2002).


\bibitem{Weiss}
U. Weiss: Quantum dissipative systems. 2nd edition, World Scientific, Singapore, 1999.



\bibitem{BDSW}
C.H. Bennett, D.P. Divincenzo, J.A. Smolin, W.K. Wootters: Mixed-state entanglement and quantum error correction. {\em Phys. Rev. A}, {\bf 54}, no.5, 3824-3851 (1996).


\bibitem{Wo}
W.K. Wootters: Entanglement of Formation of an Arbitrary State of Two Qubits. {\em Phys. Rev. Lett.} {\bf 80}, no. 10, 2245-2248 (1998).


\bibitem{HZ}
J.-H. Huang, S.-Y. Zhu: Sudden death time of two-qubit entanglement in a noisy environment. {\em Optics Communications}, {\bf 281}, 2156-2159 (2008).


\bibitem{PR}
J.P. Paz, A.J. Roncaglia: Dynamics of the entanglement between two oscillators in the same environment. Preprint arXiv:0801.0464v1.


\bibitem{Braun}
D. Braun: Creation of Entanglement by Interaction with a Common Heat Bath. {\em Phys. Rev. Lett.} {\bf 89}, 277901 (2002).


\bibitem{peres}
A. Peres: Separability Criterion for Density Matrices. {\em Phys. Rev. Lett.} {\bf 77}, 1413–1415 (1996).


\bibitem{horodecki}
M. Horodecki, P. Horodecki, R. Horodecki: Separability of Mixed States: Necessary and Sufficient Conditions. {\em Physics Letters A} {\bf 223}, 1-8 (1996).



\end{thebibliography}
\end{document}